\documentstyle[12pt,epsfig]{article}

\newcommand{\beq}{\begin{equation}}
\newcommand{\feq}[1]{\label{#1} \end{equation}}
\newcommand{\beqr}{\begin{eqnarray}}
\newcommand{\feqr}{\end{eqnarray}}
\newcommand{\non}{\nonumber}
\newcommand{\noi}{\noindent}

\newcommand{\acom}[2]{\{ #1 , #2 \}}

\newcommand{\bra}[1]{\langle #1|}
\newcommand{\ket}[1]{| #1 \rangle}

\newcommand{\rf}[1]{(\ref{#1})}

\def\alp{\alpha}
\def\bet{\beta}
\def\gam{\gamma}
\def\del{\delta}
\def\eps{\epsilon}

\def\zet{\zeta}
\def\th{\theta}
\def\vthe{\vartheta}

\def\lam{\lambda}

\def\sig{\sigma}

\def\ome{\omega}

\def\Gam{\Gamma}
\def\Del{\Delta}

\def\Lam{\Lambda}

\def\cD{{\cal D}}

\def\cG{{\cal G}}

\def\cX{{\cal X}}

\def\pa{\partial}

\def\pr{^{\prime}}

\def\rar{\rightarrow}

\newcommand{\0}{\,\!}      

\newcommand{\Dslash}{D\!\!\!\!/}

\def\np#1#2#3{Nucl. Phys. {\bf{B#1}} (#2) #3}
\def\pl#1#2#3{Phys. Lett. {\bf{#1B}} (#2) #3}
\def\prl#1#2#3{Phys. Rev. Lett. {\bf{#1}} (#2) #3}
\def\prv#1#2#3{Phys. Rev. {\bf{D#1}} (#2) #3}
\def\cm#1#2#3{Comm. Math. Phys. {\bf{#1}}, (#2), #3}
\def\qg#1#2#3{Class. Quantum Grav. {\bf{#1}}, (#2), #3}
\def\npps#1#2#3{Nucl. Phys. B (Proc. Suppl.) {\bf{#1A}} (#2) #3}

\def\ijm#1#2#3{Int. J. Mod. Phys. {\bf{A#1}} (#2) #3}
\def\rmp#1#2#3{Rev. Mod. Phys. {\bf{#1}}, (#2), #3}
\begin{document}

\hoffset = -.8in

\thispagestyle{empty}

\vspace*{-2cm}
\begin{flushright}
{\sc ITP-SB}-97-58
\end{flushright}

\vspace{1cm}

\setcounter{page}{0}

\begin{center}
{\LARGE \bf Trace and chiral anomalies in string and ordinary field theory from Feynman diagrams for nonlinear sigma models
    }\\[8mm]

\large{Agapitos Hatzinikitas\footnote{e-mail: hatzinik@insti.physics.sunysb.edu},
Koenraad Schalm\footnote{e-mail:
konrad@insti.physics.sunysb.edu}}\\
\large{and}\\
\large{Peter van Nieuwenhuizen\footnote{e-mail:
vannieu@insti.physics.sunysb.edu}}\\[6mm]

{\it Institute for Theoretical Physics\\
State University of New York at Stony Brook\\
Stony Brook, NY 11794-3840, USA}\\[6mm]

{\small \bf Abstract}\\[3mm]

\parbox{6.25in}
{\small We write general one-loop anomalies of string field theory as path integrals on a torus for the corresponding nonlinear sigma model. This extends the work of Alvarez-Gaum\'e and Witten from quantum mechanics to two dimensions. Higher world-volume loops contribute in general to nontopological anomalies and a formalism to compute these is developed. We claim that (i) for general anomalies one should not use the propagator widely used in string theory but rather the one obtained by generalization from quantum mechanics, but (ii) for chiral anomalies both propagators give the same result. As a check of this claim in a simpler model we compute trace anomalies in quantum mechanics. The propagator with a center-of-mass zero mode indeed does not give the correct result for the trace anomaly while the propagator for fluctuations $q^i (\tau)$ satisfying $q^i (\tau = -1) = q^i (\tau = 0) = 0$ yields in $d=2$ and $d=4$ dimensions the correct results from two- and three-loop graphs. 

We then return to heterotic string theory and  calculate the contributions to the anomaly from the different spin structures for $d=2$. We obtain agreement with the work of Pilch, Schellekens and Warner and that of Li in the sector with spacetime fermions. In the other sectors, where no explicit computations have been performed in the past and for which one needs higher loops,  we find a genuine divergence, whose interpretation is unclear to us. We discuss whether or not this leads to a new anomaly.}
\end{center}
\vfill

\newpage


\section{Introduction}

\setcounter{footnote}{0}

Anomalies of quantum field theories appear in the path integral approach according to Fujikawa \cite{fuji} as regularized traces of the Jacobians of the symmetry transformations whose anomalies are to be evaluated. For example, the chiral anomaly due to a spin-$\frac{1}{2}$ loop is given by ${\rm Tr} (\gam_5 \exp (-\frac{\beta}{\hbar}\Dslash \Dslash))$ for $\beta$ tending to zero, and $\Dslash\Dslash$ is the regulator where $\Dslash$ is the Dirac operator for the spin-$\frac{1}{2}$ field in the loop. A general algorithm to construct consistent regulators exists \cite{troost}. More than a decade ago, Alvarez-Gaum\'e and Witten proposed to evaluate such traces by using quantum mechanics \cite{witten}. The operators $\gam_5$, $D_{\mu}$, $x^{\mu}$, $\gam^{\mu}$ were represented by operators of a corresponding quantum mechanical model, and by turning these operator expressions into path integrals, one finds that anomalies of quantum field theories can be written in terms of Feynman diagrams for certain (nonlinear or linear) sigma models on the worldline. In particular, they computed chiral anomalies in arbitrary dimensions due to loops with spin-$\frac{1}{2}$, spin-$\frac{3}{2}$ or selfdual antisymmetric tensor fields. Their work was extended to trace anomalies in field theory by Bastianelli and van Nieuwenhuizen, who found that one needs higher loops on the worldline for these cases \cite{baspvn}. However, these authors used mode regularization, a scheme widely used at the time, and subsequent work by de Boer et. al. \cite{kostas} showed that mode regularization in general yields incorrect results for Einstein invariant Hamiltonians, although chiral anomalies come out correctly. In this article we have applied the regularization method of \cite{kostas} to the calculation of trace anomalies, following the set-up of \cite{baspvn}, and indeed find the correct results (correcting an error in \cite{baspvn}).

The main focus of this article is to extend this program from quantum mechanics to two dimensional models. It is assumed that the corresponding two dimensional traces correspond to anomalies of string field theories, but since not much is known about the latter, this is just an assumption. As has been explained before, calculations of anomalies of field theories using quantum mechanics are finite after adding ``Lee-Yang ghosts'' \cite{baspvn,lee} (ghosts due to integrating out momenta from $g^{ij}(x) p_i p_j$). Loops of a two dimensional field theory are in general divergent and anticipating that the anomalies are still finite, regularization is expected to play a role. Of course, string (field) theory is all-loop finite, hence it should not contain any anomalies at all, but the Fujikawa-like computation we are going to develop will only yield part of the anomaly (the part without the Green-Schwarz counterterm).

In a series of papers Schellekens and Warner alone \cite{warner} and in collaboration with Pilch \cite{pilch} and Lerche and Nilsson \cite{lerche} studied the one-loop chiral anomaly for the heterotic string, and the authors of \cite{warner} and Li \cite{keke} did the same for the type-II string. They assumed that this yields the chiral anomaly of a corresponding (unknown) string field theory. As generalization of the matrix $\gam_5$ they took the GSO projection operator, but anticipating that chiral anomalies come only from massless fermions the authors restricted their attention to a specific spin structure on the torus, namely the one which leads to spacetime fermions. They found that the chiral anomaly involved double sums which were only conditionally convergent. By imposing holomorphicity in the modular parameter (``world-sheet chirality'', see our appendix) the ambiguity was fixed and a finite result was obtained. The anomaly factorized and in the field theory it could then be removed by the usual Green-Schwarz mechanism. 

In this paper we explicitly evaluate the contributions to the chiral and other anomalies from all other sectors. Higher loop graphs now contribute, and we find a divergent result. This is contrary to expectation, and we discuss possible mechanisms to reduce the divergences, but whether a finite result (and hence a new anomaly) remains, requires substantial further work and insight. As the unique string field whose Jacobian we compute should contain all the information on the consistency of the theory one would in principle expect that all contributions from the other sectors vanish as well as that the anomaly calculation should reproduce the Green-Schwarz counterterms in the sector with spacetime fermions. The Fujikawa approach, however, uses the regulation prescription $\bet \rar 0$, which corresponds to analyzing the polygon graphs in the field theory limit \cite{pilch,lerche,gross}. This suggests that the naive generalization of the calculation from particle to string is incomplete and that there might exist an at present unknown extension of the Fujikawa approach, which includes the contribution of the Green-Schwarz counterterm, but we have no clear idea how to construct it. Perhaps a further Jacobian for the antisymmetric tensor fields is needed. We only note that a full first-quantized string calculation produces both the contributions of the hexagon diagram and the Green-Schwarz counterterm \cite{lerche,gross}.
    
We have developed a formalism that goes beyond one-loop determinants, and which is a direct generalization of a similar scheme constructed in quantum mechanics. We began by considering periodic or antiperiodic fluctuations in the $\sig$ and $\tau$ directions, and constructed the corresponding propagators as well-defined expressions in terms of theta functions. For the computations of effective actions this is the standard approach \cite{dhoker}. Anomalies are then given by one- or higher-loop Feynman graphs on the torus. Since the propagators are translationally invariant, one could Fourier transform these and apply dimensional regularization\footnote{For quantum mechanical models dimensional regularization is ambiguous because in expressions like $q_1 q_2 q_3 q_4$ it is not clear whether one should replace them in $n$ dimensions by $q_1\cdot q_2 \,\,\, q_3\cdot q_4$ or $q_1\cdot q_3 \,\,\, q_2\cdot q_4$ or any other invariant.}. 

However, when as a check we evaluated the trace anomalies of the $d=2$ and $d=4$ dimensional field theories by evaluating two- and three-loop graphs on the worldline using Strassler's propagator \cite{strass}, which is the one-dimensional analogue of the one used in string theory, we found an incorrect result. On the other hand for the computation of chiral anomalies it gave the correct answer. The difference of both approaches has to do with the boundary conditions on the quantum fluctuations. Decomposing $x^i(\tau) = x_0^i + q^i(\tau)$ we performed the calculation in two different ways: once by requiring $q^i(\tau)$ to vanish at the endpoints, and once by using periodic $q^i(\tau)$ which are orthogonal to the constant function on the interval $[-1,0]$. The former approach has been used extensively in \cite{kostas} to compute various anomalies in field theories, whereas the latter has been pioneered by Strassler \cite{strass} and used by him and others \cite{schubert} to compute tree graphs and loops in quantum field theories from quantum mechanics. Calculations of effective actions by way of quantum mechanics give the same answers for both methods \cite{schubert}. We present our higher loop calculations for the trace anomaly below, but first propose here an explanation of this discrepancy.

When one computes a trace of the transition matrix $\int \prod_{i=1}^d dx_0^i bra{x_0^i}\exp(-\frac{\beta}{\hbar} \hat{H})\ket{x_0^i}$, one inserts complete sets of states to turn this operator expression into a path integral. However, the combination $\int \prod_{i=1}^d dx_0^i \ket{x_0^i}\bra{x_0^i}$ is also a complete set of states, hence one can view the trace either as a problem on the line segment $[-\beta,0]$ or on a circle. In the former case one uses a complete set of functions only satisfying $x^i(\tau=-\bet) = x^i(\tau=0)$ but not $\frac{d}{d\tau}x^i (\tau=-\bet) = \frac{d}{d\tau}x^i (\tau=0)$, whereas in the latter case one uses periodic functions. Both methods should give the same answer for the partition function, and they indeed do so.

Evaluation of anomalies involves expressions such as $\int \prod_{i=1}^d dx_0^i f(x_0^i) \bra{x_0^i}\exp(-\frac{\beta}{\hbar} \hat{H})\ket{x_0^i}$ where $f(x_0^i)$ is due to the Jacobian, and now the point $x_0^i$ is special. One may view the function $f(x_0^i)$ as a vertex operator which has been inserted on the circle, thus breaking the periodicity. Schubert has shown \cite{schthe} that the answers for the effective actions differ by a total $x_0^i$ derivative thus giving the same answers after integration over $x_0^i$. Clearly, for anomalies both methods are no longer equivalent. Starting from $\bra{z^i}\exp(-\frac{\beta}{\hbar} \hat{H})\ket{y^i}$ and inserting complete sets of states, and then setting $z^i = y^i = x_0^i$, automatically selects the method with $q^i(\tau)$ vanishing at the endpoints. This is indeed what we have found to yield the correct anomalies. Strassler's method, though apparently preferable for the evaluation of effective actions, cannot be used for the evaluation of the anomalies. Rather, for anomalies one has to go back to the very definition of the operator expression and then the method with $q^i(\tau)$ vanishing at the endpoints is automatically selected. 

However, we can also take a different point of view which, at first sight, seems to lead to the opposite conclusion. Namely, exponentiating the Jacobians with a parameter $\alp$ and taking in the end the terms linear in $\alp$ one would seem to obtain a periodic action and now Strassler's propagator would be expected to yield the correct results. This is an open problem, but we suggest a resolution in the conclusion. 

Why did we get the correct answer for chiral anomalies, and what about string field theory? For chiral anomalies the relevant worldline graphs are proportional to (see eq. \rf{multi})

\beq
I_k = \int_{-1}^{0} d\tau_1 \int_{-1}^{0} d\tau_2 \ldots \int_{-1}^{0} d\tau_k \,\,\0^{\bullet}\Del^{(\rho)}(\tau_1,\tau_2)\0^{\bullet}\Del^{(\rho)}(\tau_2,\tau_3) \ldots \0^{\bullet}\Del^{(\rho)}(\tau_k,\tau_1) \,\,.
\feq{ik2}

\noi The dots denote differentiation with respect to the first variable and $\Del^{(\rho)}(\tau_1,\tau_2)$ is a generalized propagator which obeys $ \0^{\bullet \bullet} \Del (\sig, \tau) \equiv \pa_{\sig}^2 \Del^{(\rho)} (\sig, \tau) = \del(\sig-\tau) - \rho(\sig)$ for a given background charge $\rho(\sig)$ satisfying $\int_{-1}^{0} \rho(\sigma) d\sigma=1$, and equals \cite{schthe}

\beqr
\non
\Del^{(\rho)}(\tau_1,\tau_2) & = & \Del_B(\tau_1,\tau_2) - \int_{-1}^{0} d\sig \rho(\sig) \Del_B(\sig,\tau_2)  \\  && \hspace{.5in} -\int_{-1}^{0} d\sig \Del_B(\tau_1, \sig) \rho(\sig) +\int_{-1}^{0}\int_{-1}^{0} d\sig_1 d\sig_2 \rho(\sig_1) \Del_B(\sig_1,\sig_2)  \rho(\sig_2)
\label{ik3}
\feqr

\noi where $\Del_B(\tau_1,\tau_2)=\frac{1}{2} (\tau_1-\tau_2)\eps(\tau_1-\tau_2)-\frac{1}{2}(\tau_1-\tau_2)^2$. The center-of-mass propagator is obtained by setting $\rho(\sig)=1$ while the endpoint propagator corresponds to $\rho(\sig) = \del(\sig)$. In a ``link'' $\int d\tau_2 \0^{\bullet} \Del^{(\rho)}(\tau_1,\tau_2) \0^{\bullet} \Del^{(\rho)}(\tau_2,\tau_3)$, only the terms $\int d\tau_2 \0^{\bullet} \Del_B (\tau_1,\tau_2) \0^{\bullet} \Del_B (\tau_2,\tau_3)$ and  $\int d\tau_2 d\sig \0^{\bullet} \Del_B (\tau_1,\tau_2)$ \linebreak  $\0^{\bullet} \Del_B (\tau_2,\sig) \rho(\sig)$ survive due to the identity $\int_{-1}^0 d\sig \0^{\bullet} \Del_B(\sig, \tau_2) = 0$. However, all $\rho$-dependent terms in $I_k$ cancel due to the same identity. This proves that for chiral anomalies both propagators give the same result. For string field theory the same should apply to the GSO operator in the sector with spacetime fermions, but for other anomalies with functions $f(x_0^i)$ one would expect that only the method with the quantum fluctuations vanishing at the endpoints will give the correct answers. However, for the contributions to the chiral anomaly of heterotic string field theory from the sectors without spacetime fermions we shall use the center-of-mass propagator, assuming that a similar mechanism as in \rf{ik2} applies. This is current practice in string theory \cite{lerche,dhoker} but requires justification in our opinion. 

\section{Different trace anomalies from different background \\
charges}

In quantum mechanics one can split the paths $x^i(\tau)$ over which one sums in the path integral into a constant part $x_0^i$ and fluctuations $q^i(\tau)$ in many ways. These different approaches correspond to a different choice of background charges \cite{schubert}. We consider here two cases: in each case we decompose $x^i (\tau)$ into $x_0^i + q^i (\tau)$, but in the first case the fluctuations vanish at the endpoints, while in the other case the average fluctuations vanish and are taken to be periodic

\begin{description}
	\item[{\rm Case I}]: $q^i(\tau = 0) = q^i(\tau = -1) = 0 ,\,\,\,$ {\em ``endpoint approach''}
	\item[{\rm Case II}]: $\int_{-1}^0 q^i(\tau) d\tau =0 ,\,\,q^i(\tau) = q^i(\tau +1) ,\,\,\,$ {\em ``center-of-mass approach''}.
\end{description}

Following the ``center-of-mass approach'' for a real scalar field on the interval $[-1,0]$ we decompose the field $x^i(\tau)$ into a constant zero mode part $x_0^i$ and a periodic fluctuating part $q^i(\tau)$ orthogonal to the constant: $q^i(\tau) = \sum_N\; q_N^i \phi_N(\tau)$ where $\phi_N (\tau) = \left( \sqrt{2} \cos(2n\pi \tau),\right.$ $\left. \sqrt{2} \sin(2n\pi \tau) \right)$.The Green's function is given by

\beq
\Del_{cm} (\tau - \tau\pr) = \sum_{N} \frac{\phi_N (\tau) \phi_N (\tau\pr)}{\lam_N} 
\feq{delc1}

\noi where $\lam_N$ are the eigenvalues of the kinetic operator $\left( \frac{d}{d\tau} \right)^2$ and $\frac{\pa^2}{\pa\tau^2} \Del_{cm}(\tau-\tau\pr)=\delta(\tau-\tau\pr)-1$. One finds

\beq
\Del_{cm} (\tau - \tau\pr) = -2\sum_{n=1}^{\infty} \frac{\cos(2n\pi (\tau - \tau\pr))}{(2n\pi)^2} = \frac{1}{2}(\tau - \tau\pr) \epsilon(\tau - \tau\pr) - \frac{1}{2} (\tau - \tau\pr)^2 - \frac{1}{12}.
\feq{delc}

\noi This result is different from the ``endpoint Green's function'' on the interval $[-1,0]$ which satisfies $\frac{\pa^2}{\pa\tau^2}\Del_e(\tau,\tau\pr)=\delta(\tau-\tau\pr)$, 
$\Del_e(0,\tau\pr)=\Del_e(-1,\tau\pr)=\Del_e(\tau,0)=\Del_e(\tau,-1)=0$ and which is given by \cite{kostas}

\beq
\Del_e (\tau, \tau\pr) = \tau(\tau\pr +1)\theta(\tau-\tau\pr)+ \tau\pr (\tau+1)\theta(\tau\pr-\tau) = \Del_{cm} 
(\tau - \tau\pr) + \frac{1}{2}({\tau^2 +\tau\pr}^2 + \tau +\tau\pr + \frac{1}{6}).
\feq{propc}

For complex fermionic fields, the propagators in the endpoint approach were given in \cite{kostas}. One uses fermionic coherent states to evaluate $<\bar{\eta}|\exp(-\frac{\beta}{\hbar} \hat{H})|\eta>$ and decomposes the fields $\chi(\tau)$ and $\bar{\chi}(\tau)$ into constant background fields $\chi(\tau=-1)=\eta$ and $\bar{\chi}(\tau=0)=\bar{\eta}$ and quantum fluctuations $\psi(\tau)$ and $\bar{\psi}(\tau)$ which vanish at the endpoints and satisfy $<\psi(\tau) \bar{\psi}(\tau\pr)>= \theta(\tau-\tau\pr)$. In the center-of-mass approach we find for antiperiodic (AP) and  periodic (P) functions 

\beq
<\chi(\tau) \bar{\chi}({\tau\pr})>_{AP} = \sum_{n=-\infty}^{+\infty} \frac{e^{(2n+1)i\pi (\tau-\tau\pr)}}{(2n+1) i \pi } = \frac{1}{2}\epsilon(\tau - \tau\pr)
\feq{propap}

\beq
<\chi(\tau) \bar{\chi}({\tau\pr})>_{P} = {\sum_{n=-\infty}^{+\infty}}\pr   \frac{e^{2n i\pi (\tau-\tau\pr)}}{2n i \pi} = \eps(\tau-\tau\pr)- (\tau-\tau\pr).
\feq{propp}

\subsection{Chiral anomaly in the center-of-mass approach}

The chiral anomaly is given by the vacuum expectation value of $\exp(-\frac{1}{\hbar} S^{int})$ where only the vertex 
$-\frac{1}{\hbar} S^{int} = \left(-\frac{1}{\bet \hbar} \right) \frac{1}{4}\int_{-1}^{0} d\tau q^i \dot{q}^j R_{ij ab} (\omega) \lam_0^{a} \lam_0^b$ contributes \cite{witten}. The fermionic fields are periodic due to the presence of the matrix $\gam_5$ and $\lam_0^{a}$ are the zero modes of the Majorana fermions. Only $q$-loops contribute. For an $n$-dimensional field theory, the center-of-mass approach with \rf{delc} then yields 

\beqr
An(chiral) &=& \frac{(-i)^{d/2}}{(2 \pi)^{d/2}} \left(\int \prod_{i=1}^{d} dx_0^i \sqrt{g(x_0)}\right) \left(\int \prod_{a=1}^{d} d \lam_0^a \right) \non \\
&&\exp \left[\sum_{k=1}^{\infty} \left(-\frac{1}{\bet \hbar} \right)^k  \frac{(k-1)!}{k!} 2^{k-1} {\rm tr} \left(\frac{R_{ij}}{4} \right)^k (-\bet \hbar)^k I_k \right]
\label{seven}
\feqr
 
\noi where $R_{ij} = R_{ijab} \lam_0^a \lam_0^b$ and

\beq
I_k = \int_{-1}^{0} d\tau_1 \int_{-1}^{0} d\tau_2 \ldots \int_{-1}^{0} d\tau_k \, \Del_{cm}^{\bullet} (\tau_1 - \tau_2) \Del_{cm}^{\bullet} (\tau_2 - \tau_3) \ldots \Del_{cm}^{\bullet}(\tau_k - \tau_1)
\feq{multi}

\noi with $\Del_{cm}^{\bullet}(\tau_1 - \tau_2) = \frac{\pa}{\pa\tau_2}\Del_{cm}(\tau_1 - \tau_2)$. The multiple integral in \rf{multi} can be evaluated more easily than in the approach where $q^i(\tau)$ vanishes at the endpoints \cite{kostas}, by using

\beqr
\int_{-1}^{0} \Del_{cm}^{\bullet}(\tau_1, \tau_2) \Del_{cm}^{\bullet}(\tau_2, \tau_3) d\tau_2 &=& \Del_{cm} (\tau_1, \tau_3) \\
\int_{-1}^{0} \Del_{cm}(\tau_1, \tau_2) \Del_{cm}^{\bullet}(\tau_2, \tau_3) \, d\tau_2 &=& (\frac{\pa}{\pa\tau_3})^{-1}\Del_{cm} (\tau_1, \tau_3) \,\,\,\,   {\rm ,etc.}
\feqr

\noi For odd values of $k$ both ${\rm tr}(R^k)$ and $I_k$ vanish while for even $k$ one obtains the above integral which is given by

\beq
I_{2k} =  2 (-1)^{k} \sum_{l=1}^{\infty} \frac{1}{(2\pi l)^{2k}}.
\feq{five} 

\noi Substituting this result into \rf{seven} and performing first the summation over $k$ and then over $l$ yields 

\beqr
An(chiral) =  \frac{(-i)^{d/2}}{(2 \pi)^{d/2}} \left( \int \prod_{i=1}^{d} dx_0^i \sqrt{g(x_0)}\right) \left(\int \prod_{a=1}^{d} d \lam_0^a \right) \exp \left[\frac{1}{2} {\rm tr} \ln \left(\frac{\frac{R_{ij}}{4}}{\sinh \left( \frac{R_{ij}}{4}\right)}\right) \right].
\feqr

\noi This agrees with the result obtained by using the endpoint approach \cite{kostas}. It has been discussed \cite{witten}
that chiral anomalies do not depend on details of the method used to evaluate them. In the introduction we showed why.

\subsection{Trace anomalies}

We shall now evaluate the trace anomaly by both methods, but we should stress that the only propagator which is guaranteed from first principles to give the correct results is the ``endpoint propagator''. The fact that for certain calculations the center-of-mass propagator gives the correct results is not obvious at all and requires detailed analysis \cite{schubert}. Several other subtleties should be taken into account: the presence of the Christoffel terms $\Gamma\Gamma$ in the action (see\rf{intan}) and the precise rules how to evaluate products of distributions. These issues were explained in detail in \cite{kostas} where the use of the endpoint propagator was worked out. In appendix B we give the corresponding derivation for the center of mass propagator. Here we apply them to the trace anomaly. This anomaly was previously evaluated in \cite{baspvn} where no $\Gamma\Gamma$ terms were taken into account and mode regularization for products of delta and theta functions was used. It was shown in \cite{kostas} that mode regularization does not reproduce the results one gets if one lets $N$ (the number of intermediate points in the discretized path integral) tend to infinity. Rather, the delta function $\delta(\tau- \tau\pr)$ should be interpreted as a Kronecker delta function, so that for example $\int_{-1}^{0} \delta(\tau- \tau\pr) \theta(\tau- \tau\pr) \theta(\tau\pr- \tau) d\tau = \frac{1}{4}$ (and not $\frac{1}{6}$ as mode regularization would give).

In $d$ dimensions the trace (Weyl) anomaly for a real spin-0 field is given by \cite{baspvn}

\beq
An(Weyl) = Tr\left(\sig(x) e^{-\beta H}\right) = \frac{1}{(2\pi \beta \hbar)^{\frac{d}{2}}}\int \prod_{i=1}^d dx_0^i \sqrt{g(x_0^i)} \sig(x_0^i) <\exp(-\frac{1}{\hbar}S^{int})>
\feq{anom}  

\noi where

\beqr
-\frac{1}{\hbar}S^{int} = &-&\frac{1}{\beta\hbar} \int_{-1}^{0} \frac{1}{2} \left(g_{ij}(x_0+q)-g_{ij}(x_0)\right) \left(\dot{q}^i \dot{q}^j + b^i c^j + a^i a^j \right) d\tau \non \\
&-& \beta\hbar \int_{-1}^{0} \left(\frac{1}{8}R + \frac{1}{8} g^{ij} \Gamma_{ik}^{l} \Gamma_{jl}^{k} - \frac{1}{2} \xi R \right) d\tau.
\label{intan}
\feqr

\noi The Ricci curvature is defined by $R_{ij}=g^{mn} R_{imjn} = g^{mn}( \pa_i \Gam_{n ; mj} - \Gam_{in}^{k}\Gam_{k ; mj} - (i \leftrightarrow m))$ and the scalar curvature $R=g^{ij} R_{ij}$. In $d=2$ dimensions, the improvement coefficient $\xi$ ($=\frac{d-2}{d(d-1)}$) vanishes, and since propagators are proportional to $\beta\hbar$ we need tree graphs with one $\beta\hbar$ vertex or graphs with one more propagator than vertices, to cancel the factor $(\beta\hbar)^{-1}$ in the measure. Only the tree graph with a $R+\Gamma\Gamma$ vertex, and the two-loop graph with the topology of the number 8 contribute.

Using normal coordinates, we expand $g_{ij}(x_0+q)=g_{ij}- \frac{1}{3} R_{kijl}(x_0) {q}^k {q}^l +\cdots$. One finds then that the anomaly is proportional to

\beqr
&&\frac{1}{\beta\hbar}R_{iklj}(x_0) \int_{-1}^0 \frac{1}{6}<q^k q^l \left(\dot{q}^i \dot{q}^j +b^ic^j + a^ia^j\right)> d\tau - \frac{\beta\hbar}{8}R(x_0) \non \\
&=& -\frac{\beta\hbar}{6}R(x_0) \int_{-1}^{0} d\tau \left( \Del \left|\right. (\0^{\bullet}\Del^{\bullet} \left|\right.+\delta(0))-\0^{\bullet}\Del \left|\right. \,\,\0^{\bullet}\Del \left|\right. \right) - \frac{\beta\hbar}{8}R(x_0).
\label{trend}
\feqr

\noi where the vertical bar indicates that one evaluates the propagators at $\tau=\tau\pr$. The explicit $\delta(0)$ comes from the equal-time contractions of the propagator of the ghosts $a^i$, $b^i$ and $c^i$. In both approaches all terms with $\delta(\tau-\tau\pr)$ at equal time cancel. In the endpoint approach the expression in parentheses is equal to $-\tau(\tau+1)+(\tau+\frac{1}{2})^2$, and the total anomaly is $-\frac{\beta\hbar}{12}R$ which is the correct answer. In the center-of-mass approach, on the other hand, $\0^{\bullet}\Del_{cm}$ at $\tau=\tau\pr$ vanishes but $\0^{\bullet}\Del^{\bullet}_{cm}+\delta(0)=1$ at $\tau=\tau\pr$. Since $\Del_{cm}(\tau,\tau)=-\frac{1}{12}$ the total anomaly now becomes $-\frac{\beta\hbar}{9}R$. Hence, the endpoint approach yields the correct result but the center-of-mass approach does not.
 
Also in four dimensions the two methods we consider give different results. We quote here only results for the endpoint method.There are now various three-loop graphs to be evaluated. We find

\begin{figure}[h]
\hspace{1in}
\epsfxsize=1.2in
\epsfbox{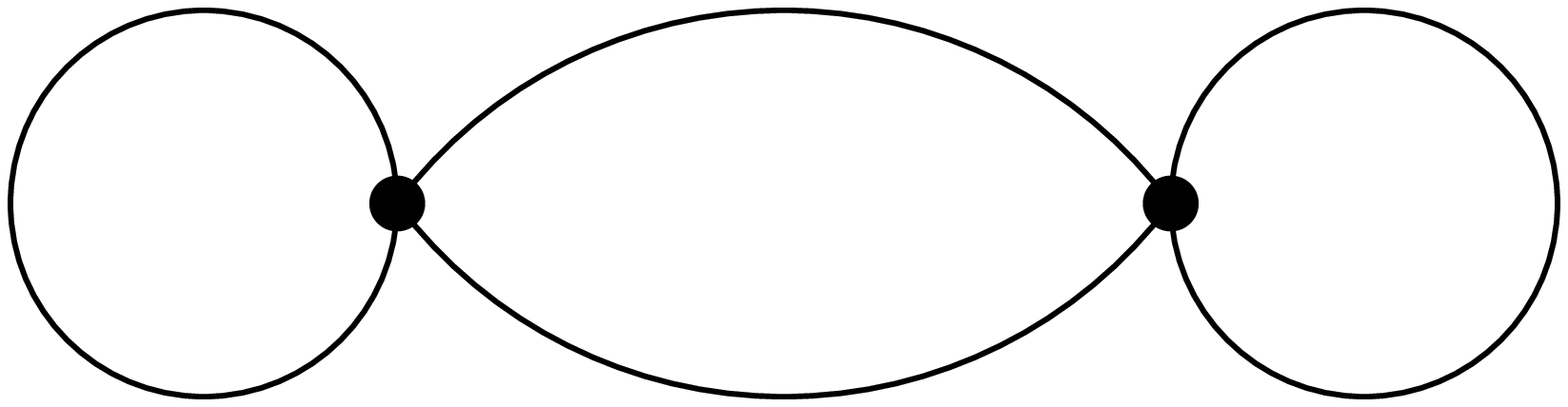}
\vspace{-.7in}
\end{figure}

\beq
\hspace{2in}
= \frac{1}{72} (-\beta\hbar)^2 \left(-\frac{1}{6} R_{ij}^2 \right)
\feq{three} 

\begin{figure}[ht]
\hspace{1.3in}
\epsfxsize=0.6in
\epsfbox{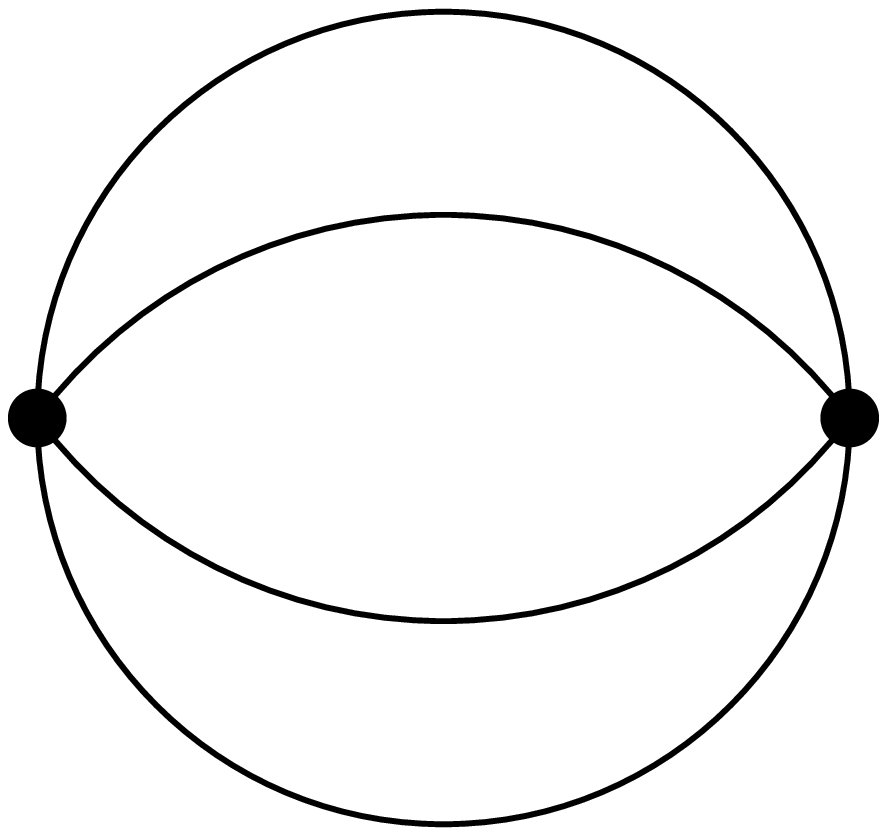}
\vspace{-.7in}
\end{figure}

\beq
\hspace{2in}
= \frac{1}{72} (-\beta\hbar)^2 \left(-\frac{1}{4} R_{ijkl}^2 \right)
\feq{eye}

\begin{figure}[ht]
\hspace{1.3in}
\epsfxsize=0.6in
\epsfbox{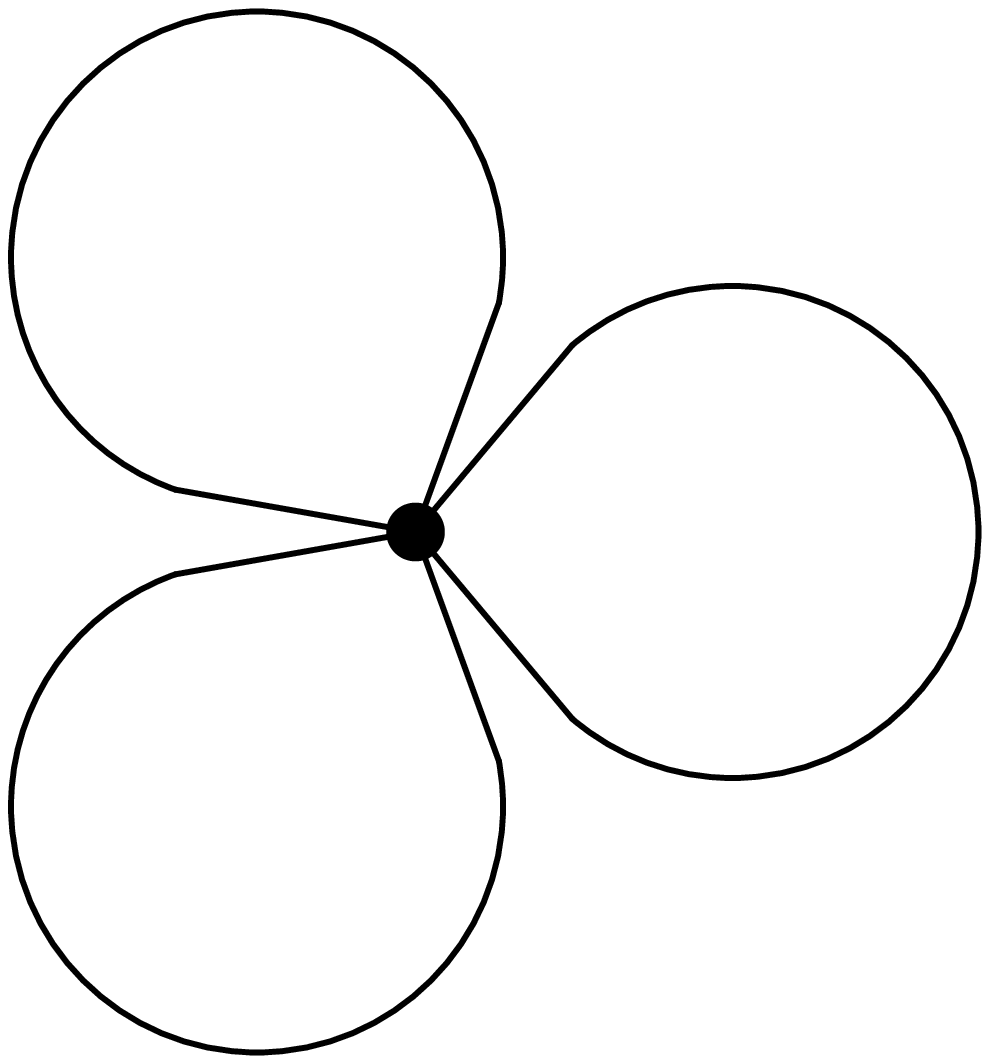}
\vspace{-.7in}
\end{figure}

\beq
\hspace{2in}
= (-\beta\hbar)^2 \left(\frac{1}{480} \nabla^2 R + \frac{1}{720} R_{ijkl}^2 + \frac{1}{1080} R_{ij}^2 \right).
\feq{glover}

\vspace{0.2in}

\noi In addition there are the contributions from the $R$, $\xi R$ and $\Gamma\Gamma$ terms (expanding both Christoffel symbols, the latter vertex yields a contribution proportional to $R_{ijkl}^2$). One finds

\begin{figure}[ht]
\hspace{1.3in}
\epsfysize=0.4in
\epsfbox{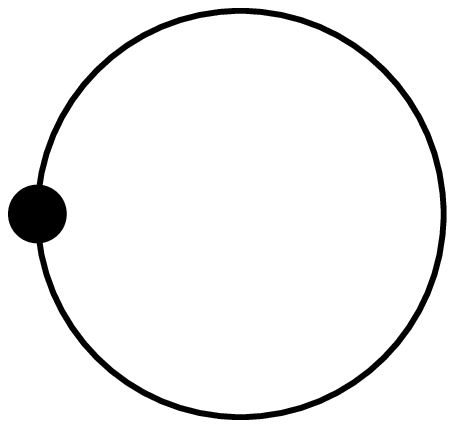}
\vspace{-.6in}

\beq
\hspace{2in}
= -\frac{1}{6} (\beta\hbar)^2 \left(\frac{1}{48} \nabla^2 R - \frac{1}{48} R_{ijkl}^2 \right).
\feq{vertex}
\end{figure}

\noi Adding all contributions, we find the correct result. 

\beq
An(Weyl) (spin \,0 , d=4) = \int \frac{d^4 x}{(2\pi)^2} \sqrt{g(x_0)} \sig(x_0) \frac{1}{720} \left(R_{ijkl}^2 - R_{ij}^2 - \nabla^2 R \right).
\feq{anom4}

We have also computed trace anomalies in $d=2$ and $d=4$ for fermions and found that in each case the endpoint method gave correct results. However, since there is now no Jacobian or measure factor, such as $\sqrt{g}$, that depends on the fermionic coordinates, it is also legitimate to use the fermionic center-of-mass propagator. It indeed results in the same answer. 

Let us now discuss whether the careful rules how to evaluate integrals over products of distributions were really necessary. One might expect that by adding integrands with the same singularities, all singularities might cancel before integration. If that were the case, any regularization scheme, provided consistently applied, would give the same answer. However, the opposite is true: whereas the singularities in the integrands with the square of $\delta(\tau -\tau\pr)$ cancel, terms linear in $\delta(\tau-\tau\pr)$ cancel almost but not completely. 
The contributions in eq. \rf{three}-\rf{glover} with $R_{ij}^2$ come from integrals whose most singular integrands are either a delta function $\del (\tau-\tau\pr)$ times a continuous function or a theta function $\th (\tau-\tau\pr)$. Any reasonable regularization scheme will give the same answers for these integrals. In the cloverleaf graph \rf{glover} one finds no $\delta(\tau -\tau\pr)$ singularities because they cancel between the $\dot{q}\dot{q}$, $bc$ and $aa$ seagull graphs. Nor are there any $\delta(\tau -\tau\pr)$ singularities due to the $\beta\hbar(R+\Gamma\Gamma)$ sector. 

However, from the eye graph \rf{eye}, proportional to $R_{ijkl}^2$,  we find two contributions involving terms with $\delta(\tau-\tau\pr)$ which combine into the following expression

\beq
\frac{1}{2} \left((\0^{\bullet}\Del^{\bullet})^2 -(\0^{\bullet\bullet}\Del)^2 \right) \Del^2 - \0^{\bullet}\Del^{\bullet}
\left(\0^{\bullet}\Del \Del^{\bullet}\right) \Del. 
\feq{delta}

\noi In the first term, terms quadratic in $\delta(\tau-\tau\pr)$ cancel (since the endpoint propagator satisfies $\0^{\bullet}\Del^{\bullet} = 1-\delta(\tau-\tau\pr)$ while $\0^{\bullet\bullet}\Del=\delta(\tau-\tau\pr)$), but a term linear in $\delta(\tau-\tau\pr)$ is left, namely $-\delta(\tau-\tau\pr) \Del^2$. In the last term we may use the identity 

\beq
\0^{\bullet}\Del \Del^{\bullet} = \Del+ \theta(\tau-\tau\pr)\theta(\tau\pr-\tau).
\feq{ident}

\setcounter{footnote}{0}

\noi Then the $\delta\Del^2$ terms also cancel, but one is left besides regular terms with an integral $I \equiv \int_{-1}^0 d\tau \int_{-1}^0 d\tau\pr \delta(\tau-\tau\pr) \theta(\tau-\tau\pr)\theta(\tau\pr-\tau) \Del(\tau,\tau\pr)$. This term is very singular, and equals $\frac{-1}{24}$ according to our regularization scheme. Other regularization schemes will in general give different results. In particular mode regularization yields $\frac{-1}{36}$. (To obtain this result use $\del (\tau-\tau\pr) = \pa_{\tau} \th (\tau-\tau\pr)$, and partially integrate to remove the delta functions.) Thus using the same action but different regularization schemes yields different results, in particular a different trace anomaly. We cannot rule out mode regularization as a consistent scheme because it may be that if one also changes the action (by adding different $\Gamma \Gamma$ -like terms) one obtains the right answers for the anomalies after all. This is actually under study. We only note that in \cite{baspvn} no $\Gamma\Gamma$ terms were taken into account since the authors used Riemann normal coordinates. At higher loop such terms will contribute, however, and redoing the calculation of \cite{baspvn} we have found an error there. Correcting this we have found that without such extra terms (or our particular choice of $\Gamma\Gamma$ terms \rf{intan}) mode regularization does not give the correct results.\footnote{We thank F. Bastianelli for collaboration in these calculations}

Our final conclusion of this section is that (i) it is only the endpoint method gives correct results for trace anomalies of scalars and fermions in $d=2$ and $d=4$ dimensions, and (ii) the noncovariant Christoffel terms and the treatment of $\delta(\tau-\tau\pr)$ as a Kronecker delta function are crucial.

\section{Anomalies for String Field Theory}

\subsection{From Regulator to Path Integral}

The classical action of the heterotic string on the Minkowski worldsheet is proportional to
  
\beqr
S_M &=& \int_{- \infty}^{\infty} dt \int_{0}^{\pi} d\sigma \left[
{1\over 2} g_{ij}(X) \partial X^i \bar{\partial} X^j +
{i\over 2} \lambda_1^a \left(
\partial \lambda_1^a 
+ \partial X^i \omega_{iab}(X) \lambda_1^b \right) \right. \non \\
&+& \left. {i\over 2} \chi_1^A \left( \bar{\partial} \chi_1^A + 
\bar{\partial} X^i A_{iAB}(X) \chi_1^B \right) + 
{1\over 4} \chi_1^A \chi_1^B \lambda_1^a \lambda_1^b F^{AB}_{ab} \right]
\feqr 

\noi where $\partial=\partial_{t} + \partial_{\sigma}$, $\bar{\partial}=\partial_{t} - \partial_{\sigma}$ and
$F^{AB}_{ab} = {\partial \over \partial X^a} A^{AB}_{b}- {\partial \over \partial X^b} A^{AB}_{a} +[A_a,A_b]^{AB}$ is the Yang-Mills curvature in a real antisymmetric representation ($A^{AB}_{a} = A^M_a (T_M)^{AB}$, $(T^{AB}_{M})^{\ast} = T^{AB}_{M}$, $T^{AB}_{M} = -T^{BA}_{M}$). This action is $N=(1,0)$ supersymmetric under 

\beq
\delta X^i = i {e^i \0_a} (X) \epsilon \lambda^a \,\, ,
\feq{susy1}

\beq
\delta \lambda^a = -e^a_i (X) \bar{\partial} X^i \epsilon - 
\delta X^i \omega_{i \,\,\,\, b}^{\,\,a} \lambda^b \,\, ,
\feq{susy2}

\beq
\delta \chi^A = -\delta X^i {A_{i} \0^A \0_B}  \chi^B \,\,.
\feq{susy3}

\noi The quantum Hamiltonian is given by

\beq
\hat{H} = {1\over 2} g^{-{1 \over 4}}\Pi_i g^{\frac{1}{2}} g^{ij} \Pi_j g^{-{1\over 4}} +
{1\over 2} \partial_{\sigma} X^i g_{ij} \partial_{\sigma} X^j - {i\over 2} \lambda^a_1 D_{\sigma} \lambda^a_1 + 
{i\over 2} \chi^A_1 {\cD}_{\sigma} \chi^A_1 - 
{1\over 4}\chi^A_1 \chi^B_1 \lambda^a_1 \lambda^b_1 F^{AB}_{ab} + k \hbar^2 R
\feq{hamil}
 
\noi where $D_{\sigma} \lambda^a = \partial_{\sigma} \lambda^a + \partial_{\sigma} X^i \omega_{i \,\,\,\, b}^{\,\,a} \lambda^b$, ${\cD}_{\sigma} \chi^A = {\partial}_{\sigma} \chi^A +{\partial}_{\sigma} X^i A_{i \,\,\,\, B}^{\,\,A} \chi^B$ and 
$\Pi_i = P_i + {1\over 2i}  \omega_{iab} \lambda^a_1 \lambda^b_1 +{1\over 2i}  A^M_i \chi_1^A (T_M)_{AB} \chi_1^B$. Below we shall encounter extra vertices not used in previous work. They are due to a careful treatment of operator orderings, but play no role for chiral anomalies. The operator orderings have been fixed in this result by imposing Einstein invariance in the same way as for the quantum mechanical case. The term $kR$ where $k$ is  a free constant to be fixed has been added since the regulator for spin-${1\over 2}$ fields contains such a term in the case of quantum field theory, so it seems possible that also such a term is present in the case of the string field theory and could be fixed by requiring that the two-loop world-sheet contribution to the chiral anomaly vanishes \cite{waldron}. 

We will use the Hamiltonian \rf{hamil} as a regulator to compute the anomaly. The closed string constraint $\hat{L}_0 = \hat{\tilde{L}}_0$ on its Hilbert space is imposed as \cite{pilch}

\beq
An = \int_{-\pi}^{\pi} d\tau_1 {\rm Tr} \left( P_{GSO} \,\, Jac \,\, e^{-\frac{\bet}{\hbar} \hat{H} + i \frac{\tau_1}{\hbar}(\hat{L}_0 - \hat{\tilde{L}}_0)} \right).
\feq{an}

\noi Here we have included a GSO projection operator to constrain the Hilbert space of the Hamiltonian \rf{hamil} to the particular string one is considering. Inserting complete sets of $X^i$ and $P_j$ eigenstates as well as fermionic coherent states for the Majorana spinors $\lam_1^a (\sig)$ and $\chi_1^A (\sig)$, one finds the continuum action which appears in the phase space path integral. It reads

\beqr
-\frac{1}{\hbar}S_E &=& \int_0^{\tau_2}dt \int_0^{\pi}d\sig \left[ \frac{i}{\hbar} P_j \dot{X}^j - \bar{\lam}_a \dot{\lam}^a - \bar{\chi}_A \dot{\chi}^A  + \bar{\Lam}_a \lam^a (\tau_2) +\bar{\cX}^A \chi_A (\tau_2) \right. \non \\ 
&-& \frac{1}{\hbar} H (P, X, \lam_1^a \rar \frac{\lam^a + \bar{\lam}^a}{\sqrt{2}}, \chi_1^A \rar \frac{\chi^A + \bar{\chi}^A}{\sqrt{2}}) \non \\
&-& \frac{\hbar}{8}\del^2(0)(R + g^{ij} \Gam^l_{ik} \Gam^k_{jl} + \frac{1}{2} g^{ij} \ome_{iab} \ome_j^{ab} + \frac{1}{2} g^{ij} A_i^M A_j^M) +\left.  {1\over \hbar} \frac{i \tau_1}{\tau_2}(L_0 - \tilde{L}_0) \right]
\label{ikosi}
\feqr

\noi where we already took the continuum limit and the last two terms in the first line are the usual boundary terms for coherent states ($\bar{\lam}_{a} = \bar{\Lam}_{a}$ and $\bar{\chi}_A = \bar{\cX}_A$ at $t = \tau_2$). The formalism we have set up does not contain anything comparable to worldsheet Einstein or supersymmetry ghosts. The reason is that we work in a flat worldsheet, so that all these ghosts are free \cite{warner}. Therefore, there are no interactions which can produce Feynman diagrams with ghosts. 

The right- and left-moving Virasoro operators $\hat{L}_0$ and  $\hat{\tilde{L}}_0$  in \rf{ikosi} are given by 

\beq 
\hat{L}_0 = {1\over 4} \left( \Pi_i - g_{ij} \pa_{\sig} X^j \right)^2 -
{i\over 2} \lam^a_1 D_{\sig} \lam^a_1 -
{1\over 8} \chi^A_1 \chi^B_1 \lam^a_1 \lam^b_1 F^{AB}_{ab}
\feq{lnull} 

\beq
{\hat{\tilde{L}}_0} = {1\over 4} \left( \Pi_i + g_{ij} \pa_{\sig} X^j \right)^2 +
{i\over 2} \chi^A_1 \cD_{\sig} \chi^A_1 -
{1\over 8} \chi^A_1 \chi^B_1 \lam^a_1 \lam^b_1 F^{AB}_{ab}.
\feq{lnulltil}

\noi Since the difference $\hat{L}_0 - \hat{\tilde{L}}_0$ 

\beq
\hat{L}_0 - \hat{\tilde{L}}_0 = -{1\over 2} \acom{\Pi_i}{\pa_{\sig} X^i} -
{i\over 2} \lam^a_1 D_{\sig} \lam^a_1 -
{i\over 2} \chi^A_1 \cD_{\sig} \chi^A_1 
\feq{pxprime}

\noi commutes with the generator of Einstein transformations

\beq
\hat{G}_E = {1\over 2 i \hbar} 
\int_0^{\pi} \left( \hat{P}_i (\sig) \cdot \xi^i ( \hat{X} (\sig)) +
\xi^i (\hat{X} (\sig)) \cdot \hat{P}_i (\sig) \right) d\sig.
\feq{gen}

\noi and is already Weyl ordered, it gives no higher-order terms.

The phase path integral can be reduced to a configuration space path integral by integrating out the momenta using the field equation for $P_i(\sig,t)$

\beq
{i\over \hbar} \pa_t X^i - {1\over \hbar} g^{ij}(X) 
\left(
P_j + {1\over 2i}\omega_{iab} \lam^a_1 \lam^a_1 +
{1 \over 2i}  A^M_i (T_M)_{AB} \chi^A_1 \chi^B_1
\right) -
{1\over \hbar}{i\tau_1\over \tau_2} \pa_{\sig} X^i = 0.
\feq{pfeq} 

\noi The final result is 

\beqr
-\frac{1}{\hbar}S_E &=& -\frac{1}{\hbar} \int_{0}^{\tau_2} dt \int_{0}^{\pi} d\sig \left[ \frac{1}{2} \sqrt{h} h^{\alp \bet} \pa_{\alp} X^i g_{ij} \pa_{\bet} X^j + \frac{1}{2} \lam_2^a \dot{\lam}_2^a + \frac{1}{2} \chi_2^A \dot{\chi}_2^A\right. \non \\
&+&  \frac{1}{2 \tau_2} \lam_1^a \left[\{ \tau_2 \pa_t - \tau_1 \pa_{\sig} - i \tau_2 \pa_{\sig} \} \lam_1^a + \{ \tau_2 \pa_t X^i - \tau_1 \pa_{\sig} X^i - i \tau_2 \pa_{\sig} X^i \} \ome_{iab} \lam_1^b \right] \non \\
&+&  \frac{1}{2 \tau_2} \chi_1^A \left[\{ \tau_2 \pa_t - \tau_1 \pa_{\sig} + i \tau_2 \pa_{\sig} \} \chi_1^A + \{ \tau_2 \pa_t X^i - \tau_1 \pa_{\sig} X^i + i \tau_2 \pa_{\sig} X^i \} A_{iAB} \chi_1^B \right]  \non \\
&-&  \left. \frac{1}{4} \chi_1^A \chi_1^B \lam_1^a \lam_1^b F_{ab}^{AB} + a \hbar^2 R + \frac{\hbar^2}{8}\del^2(0)(R + \Gam \Gam + \frac{1}{2} \ome \ome + \frac{1}{2} AA) \right].
\label{naction}
\feqr   

\noi The free spinors $\lam_2^a$ and $\chi_2^A$ are needed in a canonical treatment of the Majorana spinors $\lam_1^a$ and $\chi_1^a$ \cite{kostas}, but having obtained \rf{naction}, they can be dropped. The derivatives $\pa_{\alp} X^i$ are the usual $\pa_t X^i$ and $ \pa_{\sig} X^i$, but $\sqrt{h} h^{\alp \bet}$ is given by

\beq
\sqrt{h} h^{\alp \bet} = \frac{1}{\tau_2^2} \left( \matrix{ 	\tau_2^2 	&	-\tau_1 \tau_2 \cr
							       -\tau_1 \tau_2	&	\tau_1^2 + \tau_2^2 } \right).
\feq{metric}

\noi Rescaling $t = \tau_2 \theta$, $\lam_1 = \lam_1^\prime / \sqrt{\tau_2} $, $\chi_1 = \sqrt{\hbar} \chi_1^\prime$ and defining 

\beqr
\pa_z \equiv \pa_{\theta} - \tau \pa_{\sig} ,\,\,\pa_{\bar{z}} \equiv  \pa_{\theta} - \bar{\tau} \pa_{\sig} &;&\tau \equiv \tau_1 + i \tau_2 
\label{toot}
\feqr

\noi the action then becomes (dropping the primes on $\lam_1\pr$ and $\chi_1\pr$) 

\beqr
-\frac{1}{\hbar}S_E &=& - \int_{0}^1 d\theta  \int_{0}^{\pi} d\sig \left[ \left( \frac{1}{\tau_2 \hbar} \right) \frac{1}{2} \pa_z X^i g_{ij} \pa_{\bar{z}} X^j \right. \non \\ 
&+& \left( \frac{1}{\tau_2 \hbar} \right) \frac{1}{2} \lam_1^a D_z \lam_1^a + \frac{1}{2} \chi_1^A  \cD_{\bar{z}}  \chi_1^A  - \frac{1}{4} \chi^A \chi^B \lam_1^a \lam_1^b F_{ab}^{AB} \non \\
&+& \left.  k (\tau_2 \hbar) R + \frac{(\tau_2 \hbar)}{8}\del^2(0)(R + \Gam \Gam + \frac{1}{2} \ome \ome + \frac{1}{2} AA) \right].
\label{bel}
\feqr

\noi The terms in the last line are new w.r.t. \cite{pilch} and come from Weyl ordering of the Hamiltonian. Two equal-space commutators are needed for this, which explains the order of the term as well as the factor $\del^2(\sig)|_{\sig=0}$. They will contribute to higher loop calculations. We will discuss the associated infinities at that time.

\subsection{Propagators}

Since the action is now known, the vertices are known. It remains then to determine the propagators on the torus. Although we have argued that one should use the endpoint propagator for the quantum mechanical cases, we shall use the center-of-mass propagator for the string calculations though in general justification for this is lacking besides appeals to modular invariance. We would thus not be surprised if also for the string both approaches yield different anomalies, but it is likely that in the critical dimension the differences disappear due to the constraints of conformal invariance, which then also hold at the quantum level. The bosonic fields $X^i(\sig,\theta)$ with zero modes $x_0^i$ are decomposed as follows

\beqr
X^i(\sig,\th) =  \frac{x^i_0}{\sqrt{\pi}} + a^i_N F_N (\sig,\theta) ,\,\, \int_0^1 d\theta \int_0^\pi d\sig \frac{1}{\sqrt{\pi}} F_N = 0.
\feqr

\noi The zero modes can be thought of as the center-of-mass of the fields $X^i$, and the propagator propagates in the space of fluctuations around the center-of-mass. The propagator then becomes \cite{dhoker}

\beqr
\langle q^i (\sig,\theta) q^j (\sig\pr,\theta\pr)\rangle &=& (- \tau_2 \hbar) g^{ij}(x_0) \Del(\sig,\theta;\sig\pr,\theta\pr) \non \\
\Del(\sig,\theta;\sig\pr,\theta\pr) &=&{\sum_{m,n=-\infty}^{\,\,\, \infty \,\,\, \prime}} {1\over \lam_{m,n}} F_{m,n} (\sig,\theta) F^{\ast}_{m,n} (\sig\pr,\theta\pr)
\label{q-prop3}
\feqr 

\noi where the prime indicates that the case $m=n=0$ is to be excluded. Here $F_{m,n}(\sig,\theta)=f_n(\sig)g_m(\theta)$ where

\beqr
f_n(\sig) = \left\{ \begin{array}{ll} \sqrt{\frac{2}{\pi}} \sin 2n \sig & {\rm  for}\,\, n > 0 \\
                              \sqrt{\frac{2}{\pi}} \cos 2n \sig & {\rm for}\,\, n \leq 0    \end{array} \right. , \,\,\,\,
g_m(\th ) = \left\{ \begin{array}{ll} \sqrt{2} \sin 2m \pi \th & {\rm  for}\,\, m > 0 \\
                              \sqrt{2} \cos 2m \pi \th & {\rm  for}\,\, m \leq 0    \end{array} \right. .
\label{fg}
\feqr

\noi The functions $F_{m,n}(\sig, \th)$ are eigenfunctions of the kinetic operator $\pa_z \pa_{\bar{z}}$ with eigenvalues $\lam_{m,n} = 2i(m \pi - n \tau)\,2i( m \pi - n \bar{\tau})$ \cite{mangano}. Substituting these expressions into the general formula \rf{q-prop3}, one arrives at

\beqr
\langle q^i (\sig,\theta) q^j (\sig\pr,\theta\pr)\rangle &=& {(- \tau_2 \hbar)} \del^{ij}  {\sum_{m=0}^{\infty} \sum_{n=0}^{\infty}}\pr \frac{4}{\pi \lam_{m,n} } \cos 2n(\sig-\sig\pr) \cos 2m \pi (\theta-\theta\pr)     \non \\
&=& \frac{(- \tau_2 \hbar)}{\pi} \del^{ij} {\sum_{m,n}}\pr \frac{e^{2in(\sig-\sig\pr)+2im \pi (\theta - \theta\pr)}}{- (2 m \pi - 2 n \tau) ( 2 m \pi - 2 n \bar{\tau})} \non \\
&=& \frac{(- \tau_2 \hbar)}{\pi} \del^{ij} {\sum_{m,n}}\pr \frac{e^{\alp_{m,n}(z-z\pr)+\bet_{m,n} (\bar{z} - \bar{z}\pr)}}{\alp_{m,n} \bet_{m,n}} 
\label{q-prop2}
\feqr

\noi Here $z = \frac{i}{2\tau_2}(\sig + \bar{\tau} \th)$, $\bar{z} = \frac{-i}{2\tau_2}(\sig + \tau \th)$ and $\alp_{m,n} = 2i(m\pi - n\tau ) $, $ \bet_{m,n} = 2i(m\pi - n\bar{\tau})$. The difference between quantum mechanics and two dimensional case is manifest: in the former case a single sum over $n$ leads to $\epsilon$, $\theta$ and $\delta$ functions, but now a double sum over $m,n$ leads to Jacobi theta functions.

To obtain this propagator from the action \rf{bel} one adds and substracts a kinetic action with the metric evaluated at the point $x_0$. The term with $g_{ij}(X(\sig, \tau)) -g_{ij}(x_0)$ is left as an interaction term and after coupling to sources one integrates over all fluctuations. This gives the propagator plus an overall measure factor proportional to $\tau_2^{-d/2}$ plus the usual $1/ \eta^d (-\frac{\bar{\tau}}{\pi})\bar{\eta}^d(-\frac{\bar{\tau}}{\pi})$ terms, where $\eta(\tau)$ is the Dedekind eta function \cite{keke,gsw,gaume}. There will also be a factor proportional to $g(x_0)$ which will combine with other such factors from the derivation of propagators for the Lee-Yang ghosts to give the correct overall volume factor $\sqrt{g(x_0)}$ (see appendix B).
  
For the fermion fields we use the following complete set of functions on the torus

\beq
\phi_{m,n} (\sig ,\theta) = \frac{1}{\sqrt{\pi}} e^{2 i n \sig} e^{2 i m \pi \theta} = {1\over \sqrt{\pi}} e^{\alp_{m,n} z + \bet_{m,n} \bar{z}}
\feq{plan}

\noi where $m,n$ are either integer or half-integer depending on the boundary conditions in $\tau$ and $\sig$ respectively. We begin by constructing the propagators for the complex fields $\lam^a,\bar{\lam}_a$ and $\chi^A,\bar{\chi}_A$, but since the vertices depend only on $\lam_1 \equiv (\lam + \bar{\lam})/\sqrt{2}$ and $\chi_1 \equiv (\chi + \bar{\chi})/\sqrt{2}$, we then construct the propagators for $\lam_1^a$ and $ \chi_1^A$. The propagators for the complex (Dirac) fields are

\beqr
\langle \lam^a (\sig, \th) \bar{\lam}_b (\sig\pr \th\pr) \rangle &=& (\tau_2 \hbar) \del^a_b {\sum_{m,n=-\infty}^{\,\,\, \infty \,\,\, \prime}} \frac{1}{\alp_{m,n}} \phi_{m,n} (\sig,\theta) \phi^{\ast}_{m,n} (\sig\pr,\theta\pr) \non \\
&=& (\tau_2 \hbar) \del^a_b {\sum_{m,n=-\infty}^{\,\,\, \infty \,\,\, \prime}} \frac{1}{\alp_{m,n}} e^{\alp_{m,n}(z-z\pr)+\bet_{m,n}(\bar{z}-\bar{z}\pr)} \\
\langle \chi^A (\sig, \th) \bar{\chi}_B (\sig\pr \th\pr) \rangle &=&                \del^A_B {\sum_{m,n=-\infty}^{\,\,\, \infty \,\,\, \prime}} \frac{1}{\bet_{m,n}} \phi_{m,n} (\sig,\theta) \phi^{\ast}_{m,n} (\sig\pr,\theta\pr) \non \\
&=&  \del^A_B {\sum_{m,n=-\infty}^{\,\,\, \infty \,\,\, \prime}} \frac{1}{\bet_{m,n}} e^{\alp_{m,n}(z-z\pr)+\bet_{m,n}(\bar{z}-\bar{z}\pr)} 
\feqr

\noi The propagators for the real (Majorana) fermions then read 

\beqr
\langle \lam^a_1 (\sig, \th) \lam_1^b (\sig\pr, \th\pr) \rangle &=& \frac{(\tau_2 \hbar)}{\pi} \del^{ab} {\sum_{m,n=-\infty}^{\,\,\, \infty \,\,\, \prime}} \frac{1}{\alp_{m,n}} \sinh \{ \alp_{m,n} (z-z\pr) + \bet_{m,n} (\bar{z} - \bar{z}\pr) \}
\label{lamprop} \\
\langle \chi_1^A (\sig, \th) {\chi}_1^B (\sig\pr, \th\pr) \rangle &=& \frac{1}{\pi} \del^{AB} {\sum_{m,n=-\infty}^{\,\,\, \infty \,\,\, \prime}} \frac{1}{\bet_{m,n}}  \sinh \{ \alp_{m,n} (z-z\pr) + \bet_{m,n} (\bar{z} - \bar{z}\pr) \}
\label{chiprop}
\feqr

The kinetic terms of the fermion actions are already free and just yield the usual factors $\sqrt{\vthe_i(0, -\frac{\bar{\tau}}{\pi})/\eta(-\frac{\bar{\tau}}{\pi})}$ for each fermion \cite{keke,gsw,gaume}, where $\vthe_i$ are the Jacobi theta-functions. The index $i=1,2,3$ or 4 depending on the $\sig,\tau$ boundary conditions as in appendix A. The one provision is that the $\vthe_1$ term is to be considered without the zero mode. In that case there is also a final Grassman integral over the zero modes left with a factor $\tau_2^{d/2}$ from the rescaling in eq. \rf{toot}. 

With the above propagators, we have the following interactions  

\beqr
-\frac{1}{\hbar}S_E^{int} &=& - \int_{0}^1 d\theta  \int_{0}^{\pi} d\sig \left[ \left( \frac{1}{\tau_2 \hbar} \right) \frac{1}{2} (g_{ij}(x_0 + q) - g_{ij}(x_0)) (\pa_z q^i  \pa_{\bar{z}} q^j + b^i c^j + a^i a^j)  \right. \non \\
&+&  \left( \frac{1}{\tau_2 \hbar} \right) \frac{1}{2} \lam_1^a \pa_{z} q^i \ome_{iab}(x_0 +q)\lam_1^b + \frac{1}{2} \chi_1^A \pa_{\bar{z}} q^i A_{iAB}(x_0+q) \chi_1^B  \non \\
&-&  \frac{1}{4} \chi^A \chi^B \lam_1^a \lam_1^b F_{ab}^{AB}(x_0+q) + k (\tau_2 \hbar) R(x_0 +q) \non \\
&+& \left.  \frac{(\tau_2 \hbar)}{8}\del^2(0)(R + \Gam \Gam + \frac{1}{2} \ome \ome + \frac{1}{2} AA)(x_0+q) \right],
\label{sint}
\feqr

\noi and for the correct counting of the factors of $\tau_2$ we have an overall measure factor proportional to $\tau_2^{-d/2}$ or $1$ depending on whether $\lam$ has zero modes or not. 

\subsection{Chiral and trace anomalies for heterotic string field theory}

 In general the trace in \rf{an} will involve a GSO projection operator. This projection is needed in order to make the partition function modular invariant. However, also the generalised $\gamma_5$ is proportional to it. Consequently, for string field theory, there is no difference between chiral and trace anomalies. (Recall that also in supersymmetric quantum field theory the trace and chiral anomaly are proportional since they belong to the same anomaly multiplet.) The GSO projection, which leads to modular invariance, may or may not lead to a supersymmetric physical spectrum in the critical dimension. There are several modular invariant string theories not all of them supersymmetric, but our method applies to all of them. We shall use the $SO(32)$ string as an example.

Below we evaluate the trace of $\exp(-\frac{\beta}{\hbar}\hat{H})$ in each of the sectors with different spin structures. There are 4 sectors for $\lam$ and 4 sectors for $\chi$. We use the vertices in \rf{sint} and the propagators in \rf{q-prop2}, \rf{lamprop} and \rf{chiprop}. These are the building blocks from which to construct the full expression for the anomaly. In \cite{warner,pilch,lerche} only the sector where $\lam(\sig,\tau)$ is periodic in both $\sig$ and $\tau$, $\lam(++)$, was considered as this sector is the only one with spacetime fermions and a matrix $\gamma_5$. We shall evaluate the contributions in all sectors. In practice, however, this would require the evaluation of five-loop graphs for the $d=10$ string. One might instead consider a $d=4$ string where three-loop graphs contribute. We consider only the case $d=2$. 

From our perspective all sectors should contribute. A string field theory, being finite, does not have the freedom to move anomalies around by the introduction of counterterms as consistency of the theory uniquely determines the interactions to all orders. The complete anomaly is therefore the total contribution from all sectors. As mentioned, the relative weights of the sectors are determined by those linear combinations which in flat space (i.e. $g_{ij} = \eta_{ij}, A_j = 0 $) give a modular invariant partition function in $n$ dimensions. For instance the $SO(32)$ heterotic string ($d=10$) uses as projectors $\left[ (1-(-1)^{F_{\lam}})_R-(1-(-1)^{F_{\lam}})_{NS} \right]$ $\left[ (1+(-1)^{F_{\chi}})_R+(1+(-1)^{F_{\chi}})_{NS} \right]$ which results in the modular invariant combination $\left[ \lam (+-)\right.$ $ - \lam (++)$ $\left. - \lam (--) + \lam (-+) \right]$ $\left[ \chi (+-) + \chi (++) +  \chi (--) + \chi (-+) \right]$. Other strings use different combinations of these building blocks \cite{gsw}. The final expression for the anomaly of a string field theory is then obtained by adding the contributions of the different sectors accordingly. In their series of papers \cite{warner}, Schellekens and Warner evaluated the contribution to chiral anomalies from the $\lam(++)$ sector and found that they factorize if the theory is modular invariant, which meant that the anomaly could be cancelled by a Green Schwarz counterterm. Later, Gross and Mende, started from a modular invariant heterotic string and showed it was anomaly free \cite{gross}. 

\subsubsection*{The sectors with $\lam (++)$ and $ \chi (-+)$, $ \chi (+-)$, $\chi (--)$.}

\noi In this sector only one-loop graphs with the vertices $q^i \pa_{z} q^j (R_{ijab} \lam^a_0 \lam^b_0)$ and 
$\chi^A \chi^B (F_{ABab} \lam^a_0 \lam^b_0)$ contribute (the $\lam$ have zero modes $\lam_0^a$). They yield

\beqr
W [ \lam (++), q {\rm -loops}] &=& \sum_{k=1}^{\infty} \left( \frac{-1}{\tau_2 \hbar} \right)^k \frac{(k-1)!}{k!} \, 2^{k-1}\, {\rm tr} \left( \frac{R_{ijab} \lam^a_0 \lam^b_0}{4} \right)^k (- \tau_2 \hbar)^k I_k^1  \\
W [ \lam (++), \chi(i) {\rm -loops}] &=& -\sum_{k=1}^{\infty} \frac{(k-1)!}{k!}\, 2^{k-1}\, {\rm tr} \left( \frac{F_{ABab} \lam^a_0 \lam^b_0}{4} \right)^k  I_k^i \,\, {\rm for}\,\, i=2,3,4 
\feqr

\noi where $i=1$ corresponds to $\lam(++)$, $i=2$ to $\chi(-+)$, $i=3$ to $\chi(+-)$ and $i=4$ to $\chi(--)$. Furthermore

\beq
I_k^i = \int \frac{d^2 \sig_1 \ldots d^2 \sig_k}{(2 i \pi)^k} \left[ \prod_{j=1}^{k-1} {\sum_{u_j,v_j}}\pr \frac{e^{2 i u_j (\sig_j - \sig_{j+1}) + 2 i \pi v_j (\th_j - \th_{j+1})}}{v_j \pi - u_j \bar{\tau}} \right] {\sum_{u_k,v_k}}\pr \frac{e^{2 i u_k (\sig_k - \sig_{1}) + 
2 i \pi v_k (\th_k - \th_{1})}}{v_k \pi - u_k \bar{\tau}} \, ; 
\feq{ik}

\noi where $u_j,v_j$ are integer or half-integer depending on the (anti)periodicity in $\sig$ and $\th$ respectively. Due to the orthogonality of the plane waves, the integral can be easily evaluated. For the $I_k^1$ case one gets (see appendix A)

\beq
W [ \lam (++), q {\rm -loops}] 	= \sum_{k=1}^{\infty} \frac{1}{2k} {\rm tr} \left( \frac{R_{ij}}{4i \pi} \right)^k G_k (-\frac{\bar{\tau}}{\pi}) = -\frac{1}{2} {\rm tr} \ln \left[ \frac{\vthe_1 (R_{ij} / 4i \pi | -\frac{\bar{\tau}}{\pi})}{(R_{ij} / 4i \pi) \vthe_1\pr ( 0 |-\frac{\bar{\tau}}{\pi})} \right]
\feq{lam;q}
 
\noi where $G_k$ with $k \geq 3$ is the Eisenstein series given by \cite{pilch,schoen}  

\beq
G_k(\tau) = {\sum_{m,n}}\pr \frac{1}{(m+n \tau)^k}
\feq{eis}

\noi and $G_2$ is assumed to have been holomorphically regularized (see appendix A below \rf{app-8}). There is no contribution with $G_1$ since in this case the trace of the Riemann tensor vanishes. 

For the $\chi$-loops one finds the same result except that one must now sum  over the cases with $u,v$ equal to integer/half-integer, half-integer/integer and half-integer/half-integer values. The corresponding loop contributions are

\beqr
\non
W\left[ \lam(++), \chi(-+)\right] &=& -\sum_{k=1}^{\infty} \frac{1}{2k} {\rm tr} \left(\frac{F_{AB}}{4i \pi}\right)^k \left[G_k (-\frac{\bar{\tau}}{2\pi}) - G_k (-\frac{\bar{\tau}}{\pi})\right] \\
&=& +\frac{1}{2}  {\rm tr} \ln \left[ \frac{\vthe_4 ( \frac{F_{AB}}{4i\pi}| \frac{-\bar{\tau}}{\pi})}{\vthe_4 (0 | \frac{-\bar{\tau}}{\pi} )} \right]
\label{lam;chi4}
\feqr

\beqr
\non
W\left[ \lam(++), \chi(+-)\right] &=& -\sum_{k=1}^{\infty} \frac{1}{2k} {\rm tr} \left(\frac{F-{AB}}{4 i \pi}\right)^k \left[2^kG_k (-\frac{2\bar{\tau}}{\pi}) - G_k (-\frac{\bar{\tau}}{\pi})\right] \\
&=& +\frac{1}{2}  {\rm tr} \ln \left[ \frac{\vthe_2 (\frac{F_{AB}}{4i\pi}| \frac{-\bar{\tau}}{\pi})}{\vthe_2 (0 | \frac{-\bar{\tau}}{\pi} )} \right]
\label{lam;chi2}
\feqr

\beqr
\non
W\left[ \lam(++), \chi(--)\right] &=& -\sum_{k=1}^{\infty} \frac{1}{2k} {\rm tr} \left(\frac{F_{AB}}{4 i \pi}\right)^k \left[G_k (\frac{1}{2} -\frac{\bar{\tau}}{2\pi}) - G_k (1 -\frac{\bar{\tau}}{\pi})\right] \\
&=& +\frac{1}{2}  {\rm tr} \ln \left[ \frac{\vthe_3 (\frac{F_{AB}}{4i\pi}| \frac{-\bar{\tau}}{\pi})}{\vthe_3 (0 | \frac{-\bar{\tau}}{\pi} )} \right].
\label{lam;chi3}
\feqr

\noi One finds two sets of terms with $G_k$ on the right hand sides because we wrote the sums over half-integers = odd integers/2 as sums over all integers minus sums over even integers.

\subsubsection*{The sector with $\lam(++)$ and $ \chi(++)$.} 

We get the same loops as before (from the $q \pa q R \lam_0 \lam_0$ and $\chi \chi F\lam_0 \lam_0$ vertices) but also a new vertex contributes, namely the classical vertex $F \lam_0 \lam_0 \chi_0 \chi_0$ and the vertex $F \lam_0 \lam_0 \chi_0 \chi_{qu}$ with one quantum field. However, the tree graphs with $\chi$ propagators vanish. Hence, we are left with $q$-loops, $\chi$-loops and classical vertices. 

\beq
W\left[\lam(++), q {\rm-loops} \right] = -\frac{1}{2}  {\rm tr} \ln \left[ \frac{\vthe_1 ( \frac{R_{ij}}{4i\pi}|-\frac{\bar{\tau}}{\pi})}{\frac{R_{ij}}{4i\pi} \vthe_1\pr (0|-\frac{\bar{\tau}}{\pi})} \right]
\feq{lam;q2}

\beq
W\left[\lam(++), \chi (++) \right] = +\frac{1}{2} {\rm tr} \ln  \left[ \frac{\vthe_1 (\frac{F_{AB}}{4i\pi}|-\frac{\bar{\tau}}{\pi})}{\frac{F_{AB}}{4i\pi} \vthe_1\pr (0|-\frac{\bar{\tau}}{\pi})} \right]
\feq{lamchi;chi1}

\beq
W\left[\lam(++), {\rm classical} \right] = \frac{1}{4} F_{AB a b}\lam_0^a \lam_0^b \chi_0^A \chi_0^B. 
\feq{lamchi}

\subsubsection*{The sectors with $\lam (+-)$, $\lam (-+)$, $\lam (--)$ and $\chi (+-)$, $\chi (-+)$, $\chi (--)$.}

Since there are no zero modes in either the $\lam$ or the $\chi$ sectors for $d=2$ dimensions, a single $q$-loop or $\lam$-loop already yields the required factor $\tau_2 \hbar$, but any number of $\chi$-loops or $\chi$-trees is still allowed by $\tau_2 \hbar$ counting. The vertices which might contribute yield either two-loop graphs or zero-loop graphs
\beq
-\frac{1}{\tau_2 \hbar} \frac{1}{6} R_{kilj}(\Gamma) <q^k q^l (\pa q^i \bar{\pa} q^j + b^i c^j + a^i a^j)> \neq  0 ,\,\, -\frac{\tau_2 \hbar}{8} \del^2(0) R (1+8k) \neq 0.
\feq{q-vertex}

\noi Other one-vertex two-loop graphs vanish

\beqr
-\frac{1}{\tau_2 \hbar} \frac{1}{4} R_{ija b}(\omega)<\lam_1^a \lam_1^b q^i \pa q^j> &=& 0 ,\\
 -\frac{1}{4} F_{ijAB} <q^i \pa q^j \chi_1^A \chi_1^B> &=&  0 ,\\
  \frac{1}{4} F_{a b AB} <\lam_1^a \lam_1^b \chi_1^A \chi_1^B> &=& 0.
\label{lamvertex}
\feqr

\noi The two $q$-loop graph from eq. \rf{q-vertex} yields

\beqr
J &=& -\frac{1}{\tau_2 \hbar} \frac{1}{6} R (\tau_2 \hbar)^2 \int_0^{\pi} d\sig \int_{-1}^0 d\theta \left[ \Del (\sig, \theta; \sig\pr, \theta\pr) \left[ \0^{\pa}\Del^{\bar{\pa}} (\sig,\theta;\sig\pr, \theta\pr) + \0^{\pa \bar{\pa}}\Del(\sig, \theta; \sig\pr, \theta\pr) + \frac{1}{\pi}\right] \right. \non \\
&-& \left. \0^{\pa} \Del(\sig, \theta; \sig\pr, \theta\pr) \Del^{\bar{\pa}}(\sig, \theta; \sig\pr, \theta\pr)\right]_{\sig = \sig\pr , \theta = \theta\pr}
\label{trace-2}
\feqr

\noi The function $\0^{\pa}\Del$ is discontinuous at $\sig = \sig\pr$ and $\theta = \theta\pr$ and this ambiguity can be resolved by noticing that the propagator is a sum of products of two cosines and differentiating once will always lead to terms of products of a cosine and a sine which vanish at $\sig = \sig\pr$, $\theta  = \theta\pr$. We therefore put 

\beqr
\0^{\pa}\Del(\sig, \theta; \sig\pr, \theta\pr) \left|_{\stackrel{\sig=\sig\pr}{{\scriptscriptstyle \th=\th\pr}}} \right. = \0^{\bar{\pa}}\Del(\sig, \theta; \sig\pr, \theta\pr) \left|_{\stackrel{\sig=\sig\pr}{{\scriptscriptstyle \th=\th\pr}}} \right. = 0 &;
\feqr

\noi Also note that since the propagator $\Del(z-z\pr,\bar{z}-\bar{z}\pr)$ is translationally invariant, it satisfies

\beq
 \0^{\pa} \Del^{\bar{\pa}} + \0^{\pa \bar{\pa}} \Del= 0.
\feq{id3}

\noi Using these results in \rf{trace-2} we find the divergent answer

\beq
 J=(- \tau_2 \hbar) \frac{1}{6} R \int_0^1 \int_0^\pi d \th d \sig \left[ \frac{1}{\pi} \Del( \sig=\sig\pr, \th = \th\pr) \right] =(- \tau_2 \hbar) \frac{-1}{24 \pi} R \cG_2 (\tau, \bar{\tau})
\feq{rehec}

\noi where

\beq
\cG_k (\tau, \bar{\tau}) =  \sum_{n=1}^{\infty} 
\sum_{m=-\infty}^{\infty} \frac{1}{\left| m \pi - n {\tau} \right|^k}.  
\feq{Hecke}

\subsubsection*{The sectors with $\lam(+-)$, $ \lam(-+)$, $ \lam(--)$ and $\chi(++)$.} 

\noi The fields $\chi$ have now zero modes , and thus we get not only the same contributions as in the previous sectors with $\chi(+-)$,$\chi(-+)$,$\chi(--)$ but also contributions with the zero modes $\chi_0^A$. However, graphs of order $\tau_2 \hbar$ with two $\chi_0^A$ modes vanish due to the tracelessness of $F_{jiAB}$ and $ F_{abAB}$

\beq
-\frac{1}{4} F_{jiAB} \chi_0^A \chi_0^B < q^j \bar{\pa} q^i >=0, \,\, \frac{1}{4} F_{abAB} \chi_0^A \chi_0^B < \lam_1^a \lam_1^b >=0.
\feq{fours1}

\noi And also the graphs with one zero mode vanish

\beq
-\frac{1}{4} F_{jiAB} \chi_0^A < \chi_1^B q^j \bar{\pa} q^i >=0, \,\, \frac{1}{4} F_{abAB} \chi_0^A < \chi_1^B  \lam_1^a \lam_1^b >=0.
\feq{fours2}

\noi We see thus that we only get contributions in the sector which have no zero modes $\chi_0^A$, hence the $\chi_0$ Grassmann integral will vanish. Therefore this sector does not contribute for $d=2$.

\section{Discussion and Conclusions}
\label{conc}

We have presented a formalism to compute anomalies of string field theories which can also be used to compute higher loops. 
In particular, we have considered the sectors which were not considered in \cite{warner} and evaluated the traces for $d=2$ dimensions. In \rf{q-vertex} we have found new divergent contributions to the anomaly. The divergent series in \rf{eis} was made finite in \cite{warner} by evaluating the conditionally convergent series in a particular way (``holomorphic regularization'').  There does not seem to exist a similar principle to make \rf{rehec} finite, since holomorphicity is lost and the series is not conditionally convergent, being a sum of real positive terms. Nor can the contributions from the different sectors cancel because each sector yields the same result (it is due to a $qq$-loop) and the GSO projection operator can give only factors $\pm 1$, whereas there are an odd number of sectors with this contribution. Rather, it may be that the infinities in the $\Gamma\Gamma$ terms conspire with the infinity we have found in the two-loop calculation. Using Weyl ordering, the order $\hbar^2$ $\Gamma\Gamma$ term involves two canonical equal-time and equal-space commutators, leading to a multiplicative constant $\delta(\sigma)\delta(\sigma)$ at $\sigma=0$. The loop integral, however, diverges as $\int \frac{d^2 k}{k^2}$: in \rf{trace-2}, the singularity comes from the seagull graph with $\Del(\sig=\sig\pr, \th=\th\pr)$. One is then faced with the problem of showing that these two divergencies are proportional. This seems highly unlikely at first sight, since it is only the double derivative of the integral which is proportional to $\delta^{2}(x-y)$, but we intend to study the issue further. More likely, it could be that renormalization of the non-linear sigma model is needed and produces counterterms which contribute to the anomaly as well, in such a way as to make the latter finite. One would first have to use a symmetry principle to fix the finite parts of these counterterms, before adding all (nonfinite) contributions of the remaining sectors to decide whether or not there is a remaining anomaly. On the other hand, the renormalization of an ordinary quantum field theory does not automatically make its composite operators finite as well, hence it is not obvious (to us) that renormalization of the regulator field theory will make the anomaly finite. 

The moment one has solved the problems of the infinities, one could use our formalism to evaluate five-loop graphs which would yield the anomaly for the critical dimension $d=10$ in these sectors. In \cite{warner} it was argued that these sectors do not contain anomalies; this could be checked. It would be simpler to consider a string theory in for example four dimensions, where one would only need to evaluate three-loop graphs. If the anomaly remains divergent, however,  it would seem to suggest that we are dealing with a string field theory which is at the classical level not invariant under the transformations whose Jacobian we have equated to the GSO projection operator.

In our computation of trace anomalies using quantum mechanics we have followed the approach of \cite{kostas} in which one uses a Weyl-ordered Hamiltonian. The precise rules how to deal with products of distributions are then derived from the (finite and unambigous) discretized results by letting $N \rar \infty$. Keeping the same Hamiltonian (with the familiar $\frac{\hbar^2}{8} (R + g^{ij} \Gamma^{l}_{ik}\Gamma^{k}_{jl}$ term) but using mode regularization, leads to incorrect results as we have shown for the trace anomaly in $d=4$. However, one cannot exclude the possibility that different $\hbar$ and $\hbar^2$ counterterms in the Hamiltonian may restore mode regularization as a bona fide scheme. In fact if this is true it might hint at the existence of a consistent, although not yet known, completely covariant regularization scheme which yields the correct results.

Another issue which one should treat with care is the center-of-mass approach of Strassler, used by Schubert and others. We have shown that it yields incorrect results for field theory anomalies, but it is known to be a very efficient method to obtain correct results for effective actions. The proofs that this method gives the correct effective action have so far only be given for flat spacetime, whereas we are working in curved spacetimes. We have proposed an argument why anomalies come out incorrectly in the center-of-mass scheme (we argued that the Jacobian could be used as a vertex operator which breaks cyclic symmetry), but the argument is incomplete and needs further study. As the proof of Schubert \cite{schubert} that effective lagrangians calculated with the different schemes differ by a total derivative, seems to hold for curved spavetime as well, it might be the remaining $\sqrt{g(x_0)}$ spoils the day. Using a direct calculation, we explained why the chiral anomaly comes out correctly.

To conclude this article we now tackle the most important question which we have left unanswered till the end: is there a new anomaly from the sectors previously not considered? If the issue of divergencies will be solved, we anticipate that in the critical dimension $d=10$, the chiral and trace anomalies will not be independent (since then space-time supersymmetry is preserved and relates them). Any (if any) extra contributions to the anomaly from these sectors ought to be canceled by the supersymmetric associates of the Green-Schwarz counterterm, and we do not expect a new anomaly.

\vskip 1cm

{\bf Acknowledgments}
\nopagebreak

\noi We thank C. Hull, K. Pilch, C. Schubert, M. Strassler, N. Warner and in particular F. Bastianelli, J. de Boer and A. Schellekens for useful discussions and correspondence.


\section*{Appendix A}

\setcounter{footnote}{0}

 In \rf{lam;q}\rf{lam;chi4}\rf{lam;chi2}\rf{lam;chi3} we encountered particular sums of Eisenstein series which are related to the Jacobi $\vthe$ functions. For completeness we give here a derivation of these relations. References are \cite{warner,schoen,gaume}. We start with a few definitions.

The generalized $\vthe$ function is defined to be

\beq
\vthe \left[\matrix{ \alp \cr \bet} \right](\nu|\tau) = \sum_{n=-\infty}^{\infty} e^{[i \pi (n+\alp)^2 \tau + 2 i \pi (n+\alp)(\nu + \bet)]}\,\, .
\feq{app-1}

\noi It is related to the Jacobi $\vthe$ functions by \footnote{Note from the definition \rf{app-1} that the $\vthe$ function actually only depends on the combination $(\nu + \bet)$. Note also the minus sign in the definition of $\vthe_1$.}  

\beqr
\vthe \left[\matrix{ 0 \cr 0 }\right](\nu|\tau) &=& \vthe_3(\nu|\tau)\,\, ,\\
\vthe \left[\matrix{ 0 \cr \frac{1}{2} }\right](\nu|\tau) &=& \vthe_4(\nu|\tau)\,\, , \\
\vthe \left[\matrix{ \frac{1}{2} \cr 0 }\right](\nu|\tau) &=& \vthe_2(\nu|\tau) \,\, ,\\
\vthe \left[\matrix{ \frac{1}{2} \cr \frac{1}{2}} \right](\nu|\tau) &=& -\vthe_1(\nu|\tau) \,\, .
\feqr

\noi Using Jacobi's triple product formula we can write the $\vthe$ functions in an infinite product representation

\beq
\vthe \left[\matrix{ \alp \cr \bet }\right](\nu|\tau) = q^{-\frac{1}{24}} q^{\frac{\alp^2}{2}} e^{2 i \pi \alp (\nu + \bet)} \eta(\tau) \prod_{n=1}^{\infty} (1+q^{\alp} e^{2 i \pi  (\nu + \bet)} q^{n-1/2})(1+q^{-\alp} e^{-2 i \pi  (\nu + \bet)} q^{n-1/2})\,\, ,
\feq{app-2}

\noi where $q = e^{2 i \pi \tau}$ and $\eta(\tau)$ is the Dedekind $\eta$ function 

\beq
\eta(\tau)= e^{\frac{i \pi \tau}{12}} \prod_{n=1}^{\infty} (1-e^{2 i \pi n \tau })\,\,.
\feq{app-3}

\noi It is related to the $\vthe$ functions by the identities

\beq
\eta^3(\tau) =  \frac{1}{2\pi} \pa_{\nu} \vthe_1 (\nu|\tau) \left|_{\nu=0} \right. = \frac{1}{2} \vthe_2 (0|\tau) \vthe_3 (0|\tau) \vthe_4  (0|\tau)\,\, .
\feq{app-4}

We would like to find an expression for the sum over the Eisenstein series \rf{lam;q}

\beq
\sum_{k=1}^{\infty}\frac{1}{k} z^k G_k(\tau)\,\, .
\feq{app-5}

\noi The Eisenstein series is convergent for $k \geq 3$ and vanishes for $k$ odd. The term with $k=1$ we will assume to be absent or zero, while for $k=2$ the series is conditionally convergent and we will assume that it is given in the particular order of summation as will be used below. This is what is meant by ``holomorphic regularization''.\footnote{Evaluation of the Eisenstein series $G_2$ may also be approached by considering $\tilde{G}_2 = \lim_{k\rar 0} \sum_{m,n}\pr 1/(m+n\tau)^2 \cdot 1/|m+n\tau|^k$. Using complex analysis $\tilde{G}_2$ can be shown to differ by a non-holomorphic (in $\tau$) term $-\pi/{\rm Im}(\tau)$ from the manifestly holomorphic $G_2$ as in \rf{app-8}. Hence the name ``holomorphic regularization'' \cite{warner, schoen}.} Thus we consider the sum

\beq
\sum_{k=1}^{\infty}\frac{1}{2k} z^{2k} G_{2k}(\tau) \,\, .
\feq{app-5a}

\noi Noting that the expression for $G_k (\tau)$

\beq
G_k(\tau) = {\sum_{m,n}}\pr \frac{1}{(m+n \tau)^k}
\feq{app-6}

\noi is periodic in $\tau \rar \tau+1$, we may rewrite $G_k(\tau)$ in a Fourier series

\beq
G_k(\tau) = \sum_{r=-\infty}^{\infty} c_k e^{2 i \pi k \tau } = \sum_{r=-\infty}^{\infty} c_k q^k \,\, .
\feq{app-7}

\noi To calculate the Fourier coefficients $c_k$ we use a trick. Rewriting $G_{2k} (\tau)$ as

\beq
G_{2k}(\tau) = 2 \sum_{m=1}^{\infty} \frac{1}{m^{2k}} + 2 \sum_{n=1}^{\infty} \left\{\sum_{m=-\infty}^{\infty} \frac{1}{(m+n \tau)^{2k}}\right\} \,\, .
\feq{app-8}

\noi (This is the order of summation we assume for $G_2 (\tau)$.) Then we use the identity 

\beq 
\sum_{m=-\infty}^{\infty} \frac{1}{(m+z)^{k}} = \frac{(-1)^k (2 i \pi )^k}{(k-1)!} \sum_{n=1}^{\infty} n^{k-1} e^{2 i \pi n  z } \,\, ,
\feq{app-9}  

\noi which can be derived by taking $\frac{1}{(k-1)!} \,(\frac{d}{dz})^{k-1}$ of 

\beq
\frac{1}{z} +  \sum_{m=1}^{\infty} \left\{ \frac{1}{(m+z)} - \frac{1}{(m-z)} \right\} = \frac{\pi \cos (\pi z)}{\sin (\pi z)} = i \pi - 2 i \pi \sum_{n=0}^{\infty} e^{2 i \pi n z } \,\, .
\feq{app-10}

\noi We get

\beq
G_{2k}(\tau) = 2 \sum_{m=1}^{\infty} \frac{1}{m^{2k}} + 2 \sum_{n=1}^{\infty} \frac{(-1)^{2k} (2 i \pi)^{2k}}{(2k-1)!} \sum_{r=1}^{\infty} r^{2k-1} q^{nr} \,\, .
\feq{app-11}

\noi which is often rewritten as 

\beq
G_{2k}(\tau) = 2 \zet (2k) + 2 \frac{(2 i \pi)^{2k}}{(2k-1)!}  \sum_{a=1}^{\infty} \sig_{2k-1}(a) q^{a} \,\, ,
\feq{app-12}

\noi where $\zet(x)$ is the Riemann zeta function and $\sig_k(n)$ is the sum $\sum_{d|n} d^k$ of $k$th-powers of positive divisors of $n$.

Now we use \rf{app-11} to calculate the sum

\beqr
\non
\sum_{k=1}^{\infty}\frac{1}{2k} z^{2k} G_{2k}(\tau) &=& \sum_{k=1}^{\infty} \sum_{m=1}^{\infty} \frac{2\, z^{2k}}{2 k\, m^{2k}} + \sum_{k=1}^{\infty} \sum_{n=1}^{\infty} \sum_{r=1}^{\infty} \frac{ 2\, z^{2k} (2 i \pi r)^{2k}}{2k\, (2k-1)! \,\, r} q^{nr} \\
\non
&=& - \sum_{m=1}^{\infty} \ln (1-\frac{z^{2}}{m^{2}}) + \sum_{n=1}^{\infty} \sum_{r=1}^{\infty} \frac{2( \cos (2 \pi r z) - 1)}{r} q^{nr} \\
\non
&=&  - \ln \prod_{m=1}^{\infty}(1-\frac{z^{2}}{m^{2}}) \\
\non &&- \sum_{n=1}^{\infty} \left( \ln (1 - e^{2 i \pi z} q^n) + \ln (1 - e^{-2 i \pi z} q^n) - 2 \ln (1-q^n) \right) \\
&=& -\ln \frac{ \sin \pi z}{\pi z} \prod_{n=1}^{\infty} \frac{(1 - e^{2 i \pi z} q^n)(1 - e^{-2 i \pi z} q^n)}{(1 - q^n)^2}  \,\, .
\label{appth1}
\feqr

\noi Using the product representation of the $\vthe$ functions in \rf{app-2} we can identify the last expression with

\beq
\sum_{k=1}^{\infty}\frac{1}{2k} z^{2k} G_{2k}(\tau) = - \ln \left[ \frac{\vthe_1 (z|\tau)}{z\ 2 \pi \eta^3(\tau)} \right] = - \ln \left[ \frac{\vthe_1 (z|\tau)}{z \vthe_1\pr (0|\tau)} \right] \,\,.
\feq{app-13}
   
Similarly we find for the sum \rf{lam;chi2}

\beqr
\non
\sum_{k=1}^{\infty}\frac{1}{2k} z^{2k} \left[2^{2k} G_{2k}(2\tau) - G_{2k}(\tau) \right] 
&=&  - \ln \frac{\sin (2 \pi z)}{2 \sin (\pi z)} \\
\non &&- \ln \prod_{n=1}^{\infty} \frac{(1 - e^{4 i \pi z} q^{2n})(1 - e^{-4 i \pi z} q^{2n})(1-q^n)^2}{(1 - e^{2 i \pi z} q^n)(1 - e^{-2 i \pi z} q^n)(1-q^{2n})^2}  \\
&=& -\ln { \cos( \pi z)} \prod_{n=1}^{\infty} \frac{(1 + e^{2 i \pi z} q^n)(1 + e^{-2 i \pi z} q^n)}{(1 + q^n)^2} \,\, ,
\feqr

\noi which can be identified with

\beq
\sum_{k=1}^{\infty}\frac{1}{2k} z^{2k} \left[2^{2k} G_{2k}(2\tau) - G_{2k}(\tau) \right] = - \ln \left[ \frac{ \vthe_2 (z|\tau)}{\vthe_2(0|\tau)} \right] \,\, .
\feq{app-14}

For the sum \rf{lam;chi4}

\beq
\sum_{k=1}^{\infty}\frac{1}{2k} z^{2k} \left[G_{2k}(\frac{\tau}{2}) - G_{2k}(\tau) \right] = \sum_{k=1}^{\infty} \frac{z^{2k}}{2k} \left( 2 \sum_{r=1/2}^{\infty} \sum_{m=-\infty}^{\infty} \frac{1}{(m+r \tau)^{2k})} \right)
\feq{app-15}

\noi we notice that in \rf{appth1} the index $n$ of the final product is the same index used one of the original sums in the definition of $G_k(\tau)$. We can therefore immediately write

\beqr
\non
\sum_{k=1}^{\infty}\frac{1}{2k} z^{2k} \left[G_{2k}(\frac{\tau}{2}) - G_{2k}(\tau) \right] &=& -\ln \prod_{r=1/2}^{\infty} \frac{(1 - e^{2 i \pi z} q^{r})(1 - e^{-2 i \pi z} q^{r})}{(1 - q^{r})^2} \\
&=& - \ln \prod_{n=1}^{\infty} \frac{(1 - e^{2 i \pi z} q^{n-1/2})(1 - e^{-2 i \pi z} q^{n-1/2})}{(1 - q^{n-1/2})^2}
\feqr

\noi which is equal to

\beq
\sum_{k=1}^{\infty}\frac{1}{2k} z^{2k} \left[G_{2k}(\frac{\tau}{2}) - G_{2k}(\tau) \right] = - \ln \left[ \frac{ \vthe_4 (z|\tau)}{\vthe_4(0|\tau)} \right] \,\, .
\feq{app-15a}

Lastly, for the sum \rf{lam;chi3}

\beq
\sum_{k=1}^{\infty}\frac{1}{2k} z^{2k} \left[G_{2k}(\frac{\tau+1}{2}) - G_{2k}(\tau+1) \right] = - \ln \left[ \frac{ \vthe_4 (z|\tau+1)}{\vthe_4(0|\tau+1)} \right] \,\, ,
\feq{app-16}

\noi we use the modular properties of the $\vthe$ functions,

\beq
\vthe \left[\matrix{ \alp \cr \bet }\right](\nu|\tau+1) = e^{- i \pi ( \alp^2 - \alp)}  \vthe \left[\matrix{ \alp \cr \bet+\alp- 1/2 }\right](\nu|\tau) \,\, , 
\feq{app-17}

\noi to arrive at

\beq
\sum_{k=1}^{\infty}\frac{1}{2k} z^{2k} \left[G_{2k}(\frac{\tau+1}{2}) - G_{2k}(\tau+1) \right] = - \ln \left[ \frac{ \vthe_3 (z|\tau)}{\vthe_3(0|\tau)} \right] \,\,.
\feq{app-18}


\section*{Appendix B}

In this article we have claimed that using the center-of-mass propagator, one does not get the correct anomalies of field theories. In the computation thereof we used the calculus of delta and theta functions according to which $\del(\sig-\tau)$ is interpreted as a Kronecker delta \cite{kostas}. These rules, however, were originally derived in the context of propagators vanishing at the endpoints. This seems to leave open the possibility that different rules of handling the $\del(\sig-\tau)$ and $\th(\sig-\tau)$ distributions might lead to the correct results after all. It could also be that the propagator is subtly different from the naive continuum expression (This holds in particular for the constant part). In this appendix we show that this is not the case. The rules for products of distributions are the same as in the end-point case and the propagator is indeed the naive continuum one.

Our starting point is the kinetic part of the discretized action for the path integral of a quantum mechanical nonlinear sigma model \cite{kostas},

\beqr
Z_N(z,y;\beta) &=&\frac{\left[ g(z) g(y) \right]^{-1/4} }{(2 \pi \hbar \bet \eps)^{dN/2}} \int d^d x_1 \ldots d^d x_{N-1} \left(\prod_{k=1}^N g^{1/2}\left(\frac{x_k+x_{k-1}}{2}\right)\right) e^{-\frac{1}{\hbar} S^{(kin)}} , \label{aa1} \\
S^{(kin)} &=&  \frac{1}{2 \bet} \sum_{k=1}^N g_{ij}({x}_c) \frac{(x_k-x_{k-1})^i}{\eps}\frac{(x_k-x_{k-1})^j}{\eps} \;.
\feqr

\noi Here $\eps N =1$, $x_N = z$, $x_0 =y$ and  $x_c$ is some constant reference point which will be specified later. The $N$ factors $g^{1/2}(\frac{x_k+x_{k-1}}{2})$ are exponentiated with Lee-Yang ghosts and their treatment is the same as in \cite{baspvn, kostas}. For the calculation of anomalies one is interested in the trace of the above which puts $z=y=x_N=x_0$ and introduces an extra integration over $x_N$. The measure $\sqrt{g(x_N)}$ then cancels with part of the prefactor in \rf{aa1} and the $x$'s have become periodic. So, instead of decomposing $x$ into background and quantum fields, the latter vanishing at the ends, as was done in \cite{kostas}, we shall now decompose $x$ into periodic functions. To orthogonalize the action we use eigenfunctions of the kinetic operator

\beqr
\non
x_k &=&  \sum_{p=0}^{N/2} \frac{2}{\sqrt{N}} \cos \left( \frac{2kp \pi}{N} \right) r_p^c + \frac{2}{\sqrt{N}} \sin \left( \frac{2kp \pi}{N} \right) r_p^s \\
&=& C_k^{\;\;p}  r_p^c + S_k^{\;\;p} r_p^s \;\; ; \;\;\; k=1, \ldots, N\;,
\label{a-sub}
\feqr

\noi which are manifestly periodic in $k \rar k+N$. Here $N$ is assumed to be even and the number of modes equals the $N$ degrees of freedom we have in $x$. (Note that $\cos(2kp \pi/N)$ and  $\sin(2kp \pi/N)$ for $p > N/2$ are linearly dependent on the set in \rf{a-sub}.) The normalization is chosen such that

\beqr
\non
\sum_p C_k^{\;\;p} C_l^{\;\;p} &=& \del_{k,l} +\frac{2}{N} + \frac{2}{N}(-1)^{k-l} \;\;,\\
\sum_p S_k^{\;\;p} S_l^{\;\;p} &=& \del_{k,l}   \;.
\feqr

It is convenient to introduce the $N$ by $N$ matrix $T$

\beq
T_k^{\;\;q} = \left( C_k^{\;\;q} \;\; , \;\; S_k^{\;\;q-\frac{N}{2}} \right)\;\; ; \;\; q=0,\ldots,N-1.
\feq{a-2}

\noi To evaluate the Jacobian $J=|T|$ involved in the transformation to modes, we will first calculate $T^2$. It gives $\sum_k T_k^{\;\;p} T_k^{\;\;q} = \del_{p,q} A_q$, where $A_0 = A_{N/2} = 4$ and $A_q = 2$ otherwise. Hence the Jacobian equals $|T| = 2^{\frac{N}{2}+1}$.

Following \cite{kostas} we couple to midpoint sources. This is because the interaction part of the action, coming from a Weyl-ordered Hamiltonian, is naturally evaluated at these points,  

\beqr
-\frac{1}{\hbar} S^{(source)} &=& \sum_{k=1}^N F_{k-1/2,i} \frac{(x^i_k-x^i_{k-1})}{\eps} + G_{k-1/2,i} \frac{(x^i_k+x^i_{k-1})}{2} \;\;.
\feqr 

\noi We then have after the substitution \rf{a-sub} a partition function 

\beqr
\non
Z_N(\beta;F,G) &=& \frac{(2^{\frac{N}{2}+1})^d}{(2 \pi \hbar \bet \eps)^{dN/2}}    \int d^d r_0 \ldots d^d r_{N-1} \exp \left(  \sum_{q=0}^{N-1} - \frac{1}{\eps \bet \hbar} g_{ij}({x}_c) A_q (1- \cos \frac{2q \pi}{N}) r_q^i  r_q^j  \right. \\
&& \left.+  (\tilde{F}^q_i + \tilde{G}^q_i)r^i_q \right) \;,
\label{a-17}
\feqr

\noi where we have performed the sum over $k$. Furthermore $\tilde{F}^q \equiv \sum_{k=1}^N F_{k-1/2} \frac{(T_n^{\;\;q}-T_{k-1}^{\;\;q})}{\eps}$ and similarly for $\tilde{G}^q$. Note that at this moment the constant mode with $q=0$ decouples from the quadratic part. Since $\tilde{F}^{\; 0} \sim C_k^{\;\; 0} - C_{k-1}^{\;\; 0} = 0$ the zero mode decouples here as well, which leaves us with a single source term $\tilde{G}^{\; 0} r_0 = \sum_{k=1}^N G_{k-1/2} \frac{2}{\sqrt N} r_0$ for the mode $r_0$. We may therefore drop this source term $\tilde{G}^{\;0} r_0$ if at the same time we replace the interaction part of the action $S^{int}( \frac{\pa}{\pa F_{k-1/2}}, \frac{\pa}{\pa G_{k-1/2}})$ with $S^{int}( \frac{\pa}{\pa F_{k-1/2}}, \frac{2}{\sqrt N} r_0 + \frac{\pa}{\pa G_{k-1/2}})$. Accordingly we chose our reference point $x_c$ to be the zero mode $\frac{2}{\sqrt N} r_0$. Then completing squares for the other modes and integrating yields a contribution to the measure $\prod_{q=1}^{N-1} ( \pi \eps \bet \hbar)^{d/2} g^{-1/2}(x_c) (A_q (1- \cos \frac{2q \pi}{N}))^{-d/2}$. Combining this with the factors in \rf{a-17} and a factor $(\sqrt{N}/2)^d$ from changing the integration over $r_0$ to $x_c$, and using $\prod_{q=1}^{M-1} 2 (1- \cos \frac{q \pi}{M}) = M$, we get

\beqr
\non
Z_N(\beta;F,G) &=& \frac{1}{(2 \pi \hbar \bet)^{d/2}} \int d^d x_c \, g^{-\frac{N-1}{2}}(x_c) \exp \left( \frac{\eps \bet \hbar}{4} \sum_{q=1}^{N-1}  \frac{(\tilde{F}^q + \tilde{G}^q)_k g^{kl}(x_c) (\tilde{F}^q + \tilde{G}^q)_l}{A_q (1- \cos \frac{2q \pi}{N})} \right) .
\feqr

\noi Including $N$ factors $g^{1/2}(x_c)$ which come from the integrations over the Lee-Yang ghost kinetic term, the total measure of the partition function is identical to that of the endpoint case. 

 Now we can compute the propagators. Twice differentiating with respect to $F$ gives us

\beqr
\non
\langle \frac{x_k^i-x_{k-1}^i}{\eps}, \frac{x_m^j-x_{m-1}^j}{\eps} \rangle &=& \frac{\bet \hbar}{2 \eps} g^{ij}(x_c)\left\{ \sum_{q=1}^{N/2} \left( \frac{4}{\sqrt N} \sin \frac{q \pi}{N} \right)^2 \frac{\sin \left(2 q \pi (k-1/2)/N \right) \sin \left(2 q \pi (m-1/2)/N \right)}{A_q \left( 1- \cos(2 q \pi /N) \right)} \right.\\
\non && \left. + \sum_{q=1}^{N/2-1} \left( \frac{4}{\sqrt N} \sin \frac{q \pi}{N} \right)^2 \frac{\cos \left(2 q \pi (k-1/2)/N \right) \cos \left(2 q \pi (m-1/2)/N \right)}{A_q \left( 1- \cos(2 q \pi /N) \right)} \right\} \\
\non	&=& \frac{\bet \hbar}{2 \eps} g^{ij}(x_c)\left\{ \sum_{q=1}^{N/2-1} \frac{16}{N} \sin^2 \left(\frac{q \pi}{N}\right) \frac{\cos  \left(2 q \pi (m-k)/N \right)}{2 \left( 1- \cos(2 q \pi /N) \right)}  \right. \\
\non && \left.+ \frac{16}{N} \frac{\sin \left(\pi (k-1/2) \right) \sin \left(\pi (m-1/2) \right)}{8} \right\} \;.
\feqr

\noi In the first term the denominator cancels with the $\sin^2(q \pi/N)$ term and writing the trigonometric functions as exponentials we can do the sum to arrive at

\beqr
\langle \frac{x_k^i-x_{k-1}^i}{\eps}, \frac{x_m^j-x_{m-1}^j}{\eps} \rangle &=& (- \bet \hbar) g^{ij}(x_c) (1-\frac{1}{\eps} \del_{k,m}) \;.
\feqr

\noi At this point we find the first nontrivial result: the delta function $\del(\tau-\tau\pr)$ is again the discretized delta function of \cite{kostas}. 

Similarly we get

\beqr
\non \langle \frac{x_k^i-x_{k-1}^i}{\eps}, \frac{x_m^j+x_{m-1}^j}{2} \rangle &=& \frac{ \bet \hbar}{4} g^{ij} (x_c) \left\{\sum_{q=1}^{N/2-1} \frac{16}{N} \cos\left(\frac{q \pi}{N}\right) \sin\left(\frac{q \pi}{N}\right) \frac{\sin(2 q \pi (m-k)/N)}{2 \left( 1- \cos(2 q \pi /N) \right)} \right\} \\
	&=& (- \bet \hbar)  g^{ij}(x_c) ( \frac{1}{2} \th_{k,m} - \frac{1}{2} \th_{m,k} - \frac{(k-m)}{N}                )   \;\;,
\feqr

\noi where $\th_{m,k}$ is plus or minus 1 depending on whether $m>k$ or $m<k$ respectively and $\th_{m,m}=0$. To obtain this result we wrote the denominator as a square of sines, canceled one of them in the numerator, and used that the remaining $\sin(2 q \pi (m-k)/N)$ divided by $\sin(q \pi/N)$ yields a terminating series in $e^{i \pi/N}$.

Lastly, for the $\langle xx\rangle$ propagator

\beqr
\non
\langle \frac{x_k^i+x_{k-1}^i}{2}, \frac{x_m^j+x_{m-1}^j}{2} \rangle &=& \frac{ \eps \bet \hbar}{8} g^{ij} (x_c) \left\{\sum_{q=1}^{N/2-1} \frac{16}{N} \cos^2\left(\frac{q \pi}{N}\right) \frac{\cos(2 q \pi (m-k)/N)}{2 \left( 1- \cos(2 q \pi /N) \right)} \right\} \\
								     &=& (- \bet \hbar) g^{ij} (x_c) ( - \frac{C}{2N^2} - \frac{|m-k|^2}{2N^2} + \frac{|m-k|}{2N} - \frac{(1-\del_{m,k})}{4N} ) \;.
\feqr

\noi We write again the denominator as a square of sines, and $\cos(2 q \pi (m-k)/N)$ as $1-2\sin^2(q \pi (m-k)/N)$. The terms with $\sin^2(q \pi (m-k)/N)$ are evaluated as before and the term with factor one yields $C = \sum_{q=1}^{N/2-1} \frac{\cos^2(q \pi/N)}{\sin^2(q \pi/ N)} = \sum_{q=1}^{N/2-1} 1/ \sin^2(q \pi/ N) -N/2+1$. To evaluate this series we use that $\sum_{q=1}^{N/2-1} 1/\sin^2 (q \pi/ N) = 4 \sum_{q=1}^{N/4-1} 1/\sin^2 (2q \pi/ N) +2 $,  (use that $\sum_{q=1}^{N/2-1} 1/\sin^2 (q \pi/ N) = \sum_{q=1}^{N/2-1} 1/\cos^2 (q \pi/ N)$ and write the series as half the sum of both; we have assumed $N/2$ is even). Then the sum over even values of $q$ can be expressed in a sum over odd values, which implies that $\sum_{q=1}^{N/2-1} 1/\sin^2 (q \pi/ N) = \frac{4}{3} \sum_{q=1}^{N/4-1} 1/\sin^2 ((2q+1) \pi/ N) -\frac{2}{3}$. This last term can be computed by evaluating $\sum_{q=1}^{N/2-1} \sin^2 (q p \pi/ N)/\sin^2 (q \pi/ N)$ at $p=N/2$, since the numerator projects only onto the odd values of $q$. Rewriting the sines as exponentials one gets again a terminating series and one finds $\sum_{q=1}^{N/2-1} \sin^2 (q p \pi/ N)/\sin^2 (q \pi/ N)|_{p=N/2}= N^2/8 $. Hence $C=N^2/6-N/2+1/3$. At this point we find a second nontrivial result: the constant $1/12$ of \rf{delc} is indeed obtained from the discretized approach. 

We thus see that in the continuum limit $N \rar \infty$  we recover exactly the center-of-mass propagator from eq. \rf{delc} as well as its derivatives 

\beqr
\non
\langle x^i(\tau) x^j(\tau\pr) \rangle_{cm} &=& (- \bet \hbar) g^{ij} (x_c)\Del_{cm} (\tau,\tau\pr) \; ,\\   
\non
\langle \dot{x}^i(\tau) x^j(\tau\pr) \rangle_{cm} &=& (- \bet \hbar) g^{ij} (x_c) \0^{\bullet} \Del_{cm} (\tau,\tau\pr) \; ,\\   
\non
\langle \dot{x}^i(\tau) \dot{x}^j(\tau\pr) \rangle_{cm} &=& (- \bet \hbar) g^{ij} (x_c) \0^{\bullet}\Del_{cm}^{\bullet} (\tau,\tau\pr) \; , \\
\non 
\Del_{cm} (\tau,\tau\pr) &=& (\frac{1}{2}(\tau-\tau\pr) \eps(\tau-\tau\pr) -\frac{1}{2}(\tau-\tau\pr)^2 -\frac{1}{12})
\feqr

\noi with, however, the rules for products of distributions as in \cite{kostas}. 

One could do a similar derivation for center-of-mass fermions, which is somewhat easier due to the nature of Grassman integration. In the end it yields the expected propagators \rf{propap} and \rf{propp}, where delta and theta functions are again to be interpreted according to their discrete versions. 

\bibliographystyle{plain}

\end{document}